\documentclass[a4paper,12pt]{article}
\pdfoutput=1 

\usepackage{jheppub} 
\usepackage{amsmath}
\usepackage[T1]{fontenc} 
\newcommand{\cev}[1]{\reflectbox{\ensuremath{\vec{\reflectbox{\ensuremath{#1}}}}}}

\title{\boldmath Complexity and Emergence of Warped $\text{AdS}_3$ Space-time from Chiral Liouville Action }

\author[a,b]{Mahdis Ghodrati,}

\affiliation[a]{Center for Gravitation and Cosmology, College of Physical Science and Technology,\\
Yangzhou University, Yangzhou 225009, China}
\affiliation[b]{School of Aeronautics and Astronautics, Shanghai Jiao Tong University, Shanghai 200240, China}

\emailAdd{mahdisg@yzu.edu.cn}

\abstract{In this work we explore the complexity path integral optimization process for the case of warped $\text{AdS}_3$/warped $\text{CFT}_2$ correspondence. We first present the specific renormalization flow equations and analyze the differences with the case of CFT. We discuss how the ``chiral Liouville action" could replace the Liouville action as the suitable cost function for this case. Starting from the other side of the story, we also show how the deformed Liouville actions could be derived from the spacelike, timelike and null warped metrics and how the behaviors of boundary topological terms creating these metrics, versus the deformation parameter are consistent with our expectations. As the main results of this work, we develop many holographic tools for the case of warped $\text{AdS}_3$, which include the tensor network structure for the chiral warped CFTs, entangler function, surface/state correspondence, quantum circuits of Kac-Moody algebra and kinematic space of WAdS/WCFTs. In addition, we discuss how and why the path-integral complexity should be generalized and propose several other examples such as Polyakov, p-adic strings and Zabrodin actions as the more suitable cost functions to calculate the circuit complexity.}

\begin{document} 
\maketitle
\flushbottom

\section{Introduction}

Various tools for studying the connections between the information on a $d$ dimension boundary and the emergent $d+1$ dimensional bulk geometry have been developed in the holographic setups. Entanglement entropy would certainly play an essential role as it describes how information would be organized to arise an additional dimension. Therefore, the first holographic tool constructed in \cite{Ryu:2006bv} was for entanglement entropy and then its covariant versions were discussed in \cite{Hubeny:2007xt}.

 To understand better the interplay between information and geometry, mathematical tools such as kinematic space in the setup of integral geometry in holography was developed in \cite{Balasubramanian:2013lsa,Czech:2015qta,Czech:2016xec}.  Tensor network (TN) models such as multi-scale-renormalization ansatz (MERA) \cite{Swingle:2009bg} have been implemented to depict the construction of geometry out of organizing information which create a renormalization group flow in real space for various scales. Using this picture then the surface/state correspondence \cite{Miyaji:2015yva} has been proposed which could act as another framework for holography.
 
Then, holographic complexity which acts as another duality between geometry, (volume of a subregion, or action on a Wheeler-de Witt patch), and boundary information has been advanced in various works, \cite{Susskind:2014rva, Alishahiha:2015rta,Brown:2015bva}.  Another proposal for holographic computational complexity using path integral optimization procedure has been developed in \cite{Caputa:2017urj,Caputa:2017yrh, Bhattacharyya:2018wym, Caputa:2018kdj}. In this proposal, Liouville action plays the central role as for the two-dimensional CFTs, it would be the measure of computational complexity and it would be the functional needed to be minimized. Considering their idea, one would think how this procedure could work for other non-CFT quantum theories and which functional should replace the Liouville action. This question was the main motivation for the this work. 
 
 One of the best non-CFT theories one could consider to study such questions and developing various tools that have already been established for CFTs, is actually the warped conformal field theories (WCFTs) and its duality with warped AdS space times, $(\text{WAdS}_3/\text{WCFT}_2)$ \cite{Anninos:2008fx, Chen:2009hg, Detournay:2012pc, Chen:2013aza, Anninos:2013nja, Hofman:2014loa,Castro:2015csg, Jensen:2017tnb,Song:2017czq}. One reason this duality is very useful is that warped CFTs contain enough symmetries that many techniques of normal CFTs could still be applied. Also, in this setup, it has been shown that the entropy of a thermal state matches with the entropy of a warped BTZ black hole. In addition, the calculations for entanglement entropy and even subregion complexity have already been done for such solutions \cite{Ghodrati:2017roz, Auzzi:2018pbc, Auzzi:2019fnp}. Furthermore, interesting relationships between WCFTs  which show semi-local properties and other models such as SYK  with complex fermions which has maximal chaos have been found in \cite{Chaturvedi:2018uov}. Specifically, this fact would make studying the complexity of WCFTs more compelling, where studying the tensor network models of WCFTs and the behaviors of its partition function under renormalization, would also be useful to study the SYK model. There are even connections between WCFTs and cold atoms models \cite{Anninos:2008fx, Son:2008ye,Balasubramanian:2008dm} which make them interesting even from the applicational point of view.

In spite of these advantages, many schemes which reveal the connections between information and geometry have not yet been applied for any of these non-CFTs, including WCFTs.  In this work we aim to develop such programs for the $\text{WAdS}_3/\text{WCFT}_2$ scenario. 
 
We start by reviewing the warped conformal field theory, its global and local symmetry algebra and its holographic dual warped AdS geometry, in section \ref{sec:WADSWCFT}. Also, the related chiral Liouville gravity will be reviewed in section \ref{sec:disc}.

Then, in section \ref{sec:RGflowWCFT}, we postulate the framework of path integral optimization, RG flow equations and holographic computation complexity for warped CFT case. The idea comes from \cite{Caputa:2017urj}, where using the Liouville action and by optimization of the path integrals in conformal field theories, the time-slice of anti-de Sitter spacetime has been derived. Consequently, one of the aims of this work is to derive \textit{``warped AdS space"} using \textit{``chiral Liouville gravity"} with a similar optimization procedure where the background picture of tensor network is also in mind. 

The important difference here is the existence of two fields for the chiral gauge which both could run in the RG flow and the interplay between them produces a warped geometry where the chiral symmetry is broken. We specifically discuss these issues in section \ref{sec:twocutoffs}. We show how two different cutoffs are needed, one for each field, and then how this affects the form of the entangler function and consequently the path-integral complexity. Then, in sections \ref{sec:ActionWCFT}, we consider a scalar and a Weyl-fermion model of WCFTs where their specific actions have been derived in \cite{Hofman:2014loa,Jensen:2017tnb}. For these models, we explain how each parameter of the theory plays a role in the optimization procedure and therefore the path-integral computation complexity.

In section \ref{sec:DerAction}, we start from the other side. First, in section \ref{sec:DesAdS} we show how the Liouville action could be derived starting from an $\text{AdS}_3$ geometry, similar to the procedure shown in \cite{Takayanagi:2018pml}. Then, in sections \ref{sec:ChiralDerAct} and \ref{sec:Fibration},  we do the same thing for the spacelike, timelike and null warped $\text{AdS}_3$. We derive the deformed, chiral action for each metric and then we compare the role of the boundary, Chern-Simons term in constructing these geometries, which the results are shown in figure \ref{fig:diffterm}.

For the main part of this work, in section \ref{sec:toolsWarp}, we extend various holographic tools which have been discussed only for the AdS geometry in the literature. In section \ref{sec:TNWarped}, using chiral Liouville action versus Liouville gravity, we discuss the form of tensor networks and the kinds of gates which are needed to construct a warped CFT algebra and as a result the warped AdS geometry.  In \ref{sec:EEevolution}, we study the structure of entanglement entropy in warped CFTs and its evolution and the way each parameter such as the warping factor would play a role. In \ref{sec:sscorres}, we discuss the implications of surface/state correspondence of \cite{Miyaji:2015yva} for the case of $\text{WAdS}_3/\text{WCFT}_2$. Also, there we discuss the form of Fisher information metric for the case of warped AdS. In part \ref{sec:WCFTgatesKac}, similar to the work of \cite{Caputa:2018kdj}, we build the quantum circuits of Kac-Moody algebra and then we calculate its cost functions and then the circuit complexity.  In part \ref{sec:WarpedKinematic}, we calculate the Crofton form of WCFTs and discuss the geometry of kinematic space for this case.

In section \ref{sec:Polymodes}, we discuss the importance of generalizing Liouville action as the cost function for calculating complexity, to Polyakov action and then its various generalizations such as p-adic strings and Zabrodin actions.

Finally, we conclude with a discussion in section \ref{sec:Discuss}.

\section{Warped Conformal Field theory and Holography}

In this section, we first review warped conformal field theories (CFTs) and the chiral Liouville gravity. We discuss the symmetries, group algebra, operators, the metric of warped AdS, the derivation of chiral Liouville theory from Polyakov action using chiral gauge, the equations of motion and the meaning of various parameters in the holographic picture.

\subsection{$\text{WAdS}_3$/$\text{WCFT}_2$}\label{sec:WADSWCFT}

In \cite{Hofman:2011zj}, the idea of omitting one of the four global symmetries of the CFT which then could lead to a chiral, Lorentz-breaking $2d$ QFT has been studied. This in fact would lead to a theory which has only one, ``left moving, global scaling symmetry". The authors there discussed how the left conformal symmetry is still intact, while depending on the assumptions, the right translations would be enhanced either to a right conformal or a left $U(1)$ ``Kac-Moody symmetry". Here, we always consider the later case.

So the global symmetries of the warped CFT that we consider are
\begin{gather}\label{eq:symmetries}
x^- \to x^- +a, \ \ \ \ \ \ x^+ \to x^+ +b, \   \ \ \ \ \ 
x^- \to \lambda^- x^-.
\end{gather}

Therefore, these kinds of field theories only lack the the dilation symmetry of $x^+ \to \lambda^+ x^+$.  Also, instead of Brown-Henneaux boundary condition \cite{Brown:1986nw}, the correct boundary algebra of WCFTs would be the CSS boundary conditions discussed in \cite{Henneaux:2011hv, Compere:2013bya}.

The goal is to study how the lack of this dilation symmetry for the $x^+$ case, (or any other global chosen, omitted symmetry which could lead to other field theories) and also perhaps the choice of boundary conditions, would affect the emergence of spacetimes out of the boundary information and any proposed tensor network for such theories.

The connections between local and global symmetries in the boundary and bulk in the AdS/CFT setup, their exact definitions and the interplay between them have been recently studied in more details in \cite{Harlow:2018tng, Harlow:2018jwu}. Their results would help to construct the tensor network model for each of these deformed CFTs. We would also like to see how lack of this symmetry would change the optimization procedure proposed in \cite{Caputa:2017yrh, Caputa:2017urj} and how it would change the properties of computational complexity.

Along the way, we use many known results for the partition function, properties of the dual holographic gravitational theories such as chiral Liouville action, the proposed actions for WCFTs such as those found in \cite{Jensen:2017tnb}, the anomaly behaviors, and the structure of entanglement entropy and holographic complexity of warped CFTs derived in previous works  \cite{Ghodrati:2017roz,Ghodrati:2018hss}.

It worths to mention here that in fact there are various examples in real world for these field theories, such as the continuum limit of large $N$ chiral Potts model \cite{Hofman:2011zj,Cardy:1992tq} which is a spin model on a planar lattice. In this model each spin could have $n=0,..., N-1$ values and to each pair of nearest neighbor of spins $n$ and $n'$, a Boltzman weight $W(n-n')$ would be assigned. For the chiral case though $W(n-n') \ne W(n'-n)$, but as these weights satisfy the Yang-Baxter equation, the model would be integrable.  So considering these real-world systems and their applications, would make studying various quantum information properties of these models, such as holographic complexity or their models of tensor network much more interesting, even from a practical point of view.

So as the reminder, we note that the operators of WCFTs satisfy the following commutation relations on the plane \cite{Detournay:2012pc}
\begin{flalign}
i[T_\xi, T_\zeta] &=T_{\xi' \zeta-\zeta' \xi} +\frac{c}{48\pi} \int dx^- ( \xi" \zeta'-\zeta" \xi'),\nonumber\\
i[P_\chi, P_\psi] &=\frac{k}{8\pi} \int dx^- (\chi' \psi-\psi',\nonumber\\
i[T_\xi, P_\chi] &= P_{-\chi' \xi},
\end{flalign}
where the local operators
\begin{gather}
T_\xi=-\frac{1}{2\pi} \int dx^- \xi(x^-) T(x^-), \ \ \ \ \ \ P_\chi=-\frac{1}{2\pi} \int dx^- \chi(x^-) P(x^-),
\end{gather}
are the left moving stress tensor and the left moving $U(1)$ current respectively. The right moving is associated with $x^-$ and left moving with $x^+$. 

Then, with the change of coordinate $x^-=e^{i\phi}$ to go from plane to cylinder, the Kac-Moody algebra of warped CFTs could be written as \cite{Detournay:2012pc}
\begin{flalign}
[L_n,L_m] &=(n-m)L_{n+m}+\frac{c}{12} n (n-1)(n+1) \delta_{n+m},\nonumber\\
[P_n,P_m]&=\frac{k}{2} n \delta_{n+m},\nonumber\\
[L_n,P_m]&=-mP_{m+n},
\end{flalign}
where $L_n=iT_{\xi_{n+1}}$, $P_n=P_{\chi_n}$ are the operators on the cylinder and $\xi_n=(x^-)^n=e^{in\phi}$ are the test functions. Also, $c$ is the central charge of the Virasoro algebra and $k$ is the Kac-Moody level.

Considering a chiral gauge, one could consider a general metric for the background of warped CFTs as
\begin{gather}\label{eq:chiralmetricgauge}
ds^2=-e^{2 \rho (t^+, t^-) } \Big ( dt^+ dt^- - h(t^+, t^-) (dt^+)^2  \Big), \ \ \ \ \ \ \ \ \  \partial_- h=0.
\end{gather}

The symmetries acting infinitesimally on the coordinates $t^\pm$, instead of being $\delta t^+=\epsilon^+(t^+)$ and $\delta t^-= \epsilon^- (t^-)$ as for the conformal case would be $\delta t^+= \epsilon(t^+)$ and $\delta t^- =\sigma(t^+)$ for the chiral case, which at the end will lead to two non-zero Noether conserved charge $j_\epsilon^-$ and $j_\sigma^-$. 

We will explain further about this metric in the next part.

In the holographic setup, these field theories holographically are the dual to the warped $\text{AdS}_3$ geometries in the bulk. The metric of warped $\text{AdS}_3$ could be written as
\begin{gather}
g_{WADS_3}= L^2 \left ( \frac{dr^2}{r^2} -r^2 {dx^-}^2 +\alpha^2   ({dx^+} +r dx^- )^2    \right),
\end{gather}
which is a fibration over $\text{AdS}_2$ with a squashing parameter $\alpha$.  For $\alpha=1$, this is an $\text{AdS}_3$ metric.

The global isometry of these geometries is $SL(2;\mathbb{R})\times U(1)$ and with a suitable boundary condition this asymptotic symmetry group would be enhanced to the Virasoro-Kac-Moody algebra.

Also, note that the above geometries appear in the near-horizon limit of four-dimensional extremal Kerr black holes where at fixed polar angle the three-dimensional warped $\text{AdS}_3$ would appear, leading to the Kerr/CFT correspondence.

\subsection{Chiral Liouville Gravity}\label{sec:disc}
Here, we review the chiral Liouville action, the chiral gauge and its connection to WAdS/WCFT.

In a conformal gauge, any metric of a gravitational theory should satisfy
\begin{gather}\label{eq:confgauge}
g_{--}=g_{++}=0,
\end{gather}
which is dual to a conformal field theory without gravity where the component $g_{+-}$ acts as one of the fields. The classical $2d$ Liouville gravity is written in this gauge and has a residual left and right Virasoro symmetry algebra.

However, one could alternatively choose the ``chiral gauge" \cite{Compere:2013aya}
\begin{gather}
g_{--}=0,\ \ \ \ \ \partial_-(g^{+-} g_{++})=0,
\end{gather}
which on the boundary side would lead to a warped conformal field theory without gravity. The residual symmetry is a right  Virasoro Kac-Moody algebra with no left Virasoro symmetry group. The general metric which satisfy such constraints would be of the form \ref{eq:chiralmetricgauge},
\begin{gather}
ds^2=-e^{2\rho} (dt^+ dt^--h(dt^+)^2 ).
\end{gather}
Note that from the chiral gauge constraint, we have also the relation $\partial_- h=0$ .

The usual Liouville gravity, in fact, has been written starting from the nonlocal Polyakov action \cite{Compere:2013aya}
\begin{gather}
S^0_L= \frac{c}{96\pi} \int d^2x (Z \mathcal{R} -2 \lambda \sqrt{-g}),
\end{gather}
with the following definitions for the parameters,
\begin{gather}
\mathcal{R} \equiv \sqrt{-g} R,\nonumber\\
Z(x) \equiv \int d^2 x' G(x,x') \mathcal{R} (x'),\nonumber\\
\sqrt{-g} g^{ab} \nabla_a \nabla_b G(x,x') = \delta^2 (x,x').
\end{gather}

Then, by gauge fixing it, using the conformal gauge \ref{eq:confgauge}, one gets the Liouville gravity theory.

For the metric \ref{eq:chiralmetricgauge}, the Ricci density would be
\begin{gather}\label{eq:ricciwarped}
\mathcal{R}=4\partial_- \partial_+ \rho+4 h \partial^2_-\rho +[4\partial_- h \partial_- \rho+2 \partial^2 _- h]=-2\sqrt{-g} \nabla^2 \rho+[2\partial^2_ - h].
\end{gather}

The terms in the bracket will vanish only when the gauge condition $\partial_- h=0$ is imposed. For such case, considering  the chiral gauge, the scalar curvature would only depend on the gradient of the field $\rho$. This point would be important in constructing the tensor network and understanding the emergent bulk spacetimes out of these chiral theories.

Additionally, the equations of motion for $\rho$ and $h$ from the action \eqref{action1} would be \cite{Compere:2013aya}
\begin{gather}\label{eq:eqofmotion}
\partial_+ \partial_- \rho=0, \ \ \ \   \ \ \ \ \ \ \ \
\partial_- \partial_+ \rho+2h \  {\partial^2_-} \rho+{\partial^2_-} h=0.
\end{gather}

On a solution of \eqref{eq:eqofmotion}, one could get the complexity of a quantum state which is specified by the boundary conditions of $\rho$ and $h$.

To study the chiral Liouville theory in a fixed sector for the left moving zero modes, an additional term would be added to $S^0_L$ in the following from \cite{Compere:2013aya}
\begin{gather}
S_L=S^0_L+\frac{\Delta}{4\pi} \int d^x \sqrt{-g} g^{--}=S^0_L-\frac{\Delta}{2\pi} \int d^2 x h,
\end{gather}
and the equation of motion for $g^{--}$ would be $T_{--}=T^0_{--}+\frac{\Delta}{2}=0$, where $\Delta/2$ is the left-moving energy density. Now the idea is to look for the effects of this additional term on the computational complexity and the tensor network structures. Setting $t^-$ as the time, the above action could then be written as the form of relation \ref{eq:ChiralLiouvilletwo}.

In \cite{Caputa:2017yrh}, one way of evaluating computational complexity is proposed to be derived by minimizing the Liouville action, where each term, i.e, the kinetic and the potential parts, would determine the number of quantum gates in the tensor network as \cite{Caputa:2017yrh, Czech:2017ryf}
\begin{flalign}
S_L&=\frac{c}{24 \pi} \int dx \int_\epsilon^\infty dz \Big[ \underbrace{ (\partial_x \phi)^2+ (\partial_z \phi)^2}_{ \# \text{ of Isometries \includegraphics[width=4mm]{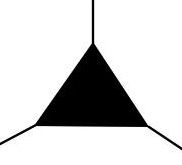} }} +\underbrace{\delta^{-2} e^{2\phi}}_{\# \text{ of Unitaries \includegraphics[width=4mm] {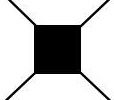} } }\Big].
\end{flalign}

Now we would like to check if one could also derive the time slice of warped $\text{AdS}_3$ from warped CFTs, by using the chiral Liouville theory , and also by considering the connections between the different terms in that action with the number of quantum gates of the tensor network structure.

Furthermore, in \cite{Czech:2017ryf},  the complexity of path integral $\mathcal{C}[\phi]$ has been characterized as the functional of $\phi(z,x)$ where varying that could determine the minimally complex preparation of the state, leading to the derivation of Einstein equations by varying complexity.  One would expect the same strategy could work here as the complexity of a path integral carried over the warped $\text{AdS}_3$ would be the chiral Liouville action.  So the equations of motion \ref{eq:eqofmotion} could be derived by varying the complexity here. For demonstrating this idea, we could also use the methods of \cite{Caputa:2017urj}.

In fact, in our previous work \cite{Ghodrati:2017roz} we proposed the following connections,
\begin{flalign}\label{eq:ChiralLiouvilletwo}
S_L&=\frac{c}{12 \pi} \int d^2 x \left( \underbrace{   \partial_{+} \rho  \partial_{-} \rho}_{\text{$\#$ of Isometries}} -\underbrace{ \frac{\Lambda}{8} e^{2\rho}}_{\# \text{ of Unitaries}} + \underbrace{ h(\partial_{-} \rho)^2 + [ \partial_- h \partial_- \rho ] -\frac{6}{c} h \Delta }_{\# \text{ of WCFTs new gate, Chiraleons}} \right ),
\end{flalign}
or one can also write this action in the following form
\begin{flalign}\label{action1}
S&=S_L^0 +\int dt^+ dt^- \left( \underbrace{\partial_+ \phi \partial_- \phi}_{\# \text{  of Isometries}} +\underbrace{h \partial_- \phi \partial_- \phi}_{\text{$\#$ of WCFTs new gate}}\color{black}-\underbrace {\frac{m^2}{4} e^{2\rho} \phi^2 }_{\# \text{ of Unitaries}} \right) .
\end{flalign}

Similar to the case of \cite{Balasubramanian:2013lsa,Czech:2015qta,Czech:2016xec}, we took the kinetic term of the chiral action to be proportional to the number of isometries, and the potential term to the number of unitaries or disentangler \cite{PhysRevLett.115.180405}. As no combinations of these gates could lead to the deforming term, then for creating this term, one would need an additional kind of quantum gates which we dub ``chiralons".

So the chiral theory here could be considered as the usual CFT deformed by a term, $h \partial_- \phi \partial_- \phi$. In this term $h$ is proportional to a right moving current and has dimension $(1,0)$, the term $(\partial_- \phi)^2$ has dimension $(0,2)$, therefore the middle term is a dimension $(1,2)$ operator. Such deforming operators which appear in WCFT and Kerr/CFT are related to the IR limit of the \textit{``dipole deformed gauge theories"} \cite{Bergman:2001rw,Bergman:2000cw}.

For constructing a tensor network structure for these theories, one could imagine several subtleties and complications. For instance the chirality of WCFTs could break the sense of locality of the network. However, it has been shown that, in fact these theories are semi-local, and therefore have enough locality to be constructed using a ``semi-local entangler function". Also, the involuted way of coupling the two fields $h$ and $\rho$ would be another issue, though considering enough constraints such as $\partial_- h=0$ would make many simplifications here, as it has been observed in the relation  for Ricci scalar \ref{eq:ricciwarped}. Also, the fermion doubling issue when a fermion theory is being put on a lattice should be in mind too which we will discuss further later.

 There are though beautiful studies of the fractal structure of gravity models as in \cite{Polyakov:1987zb, Knizhnik:1988ak} in the setup of $\gamma$-Liouville quantum gravity (LQG) surfaces, where $\gamma \in (0,2)$ which could be applied here.  One could check how this structure would be related to the emergence of spacetime, the structure of tensor network, and the complexity.  For instance one of the Kac-Moody current $j_\sigma$ mentioned here is related to one of the Knizhnik-Polyakov-Zamolodchikov (KPZ) $\text{SL}(2,\text{R})$ currents in \cite{Knizhnik:1988ak}. Further methods of discretizations of these theories could be studied.

\section{Wilsonian RG Flow for WCFTs}\label{sec:RGflowWCFT}

Now in this section, similar to the studies of \cite{Caputa:2017urj,Caputa:2017yrh, Bhattacharyya:2018wym, Caputa:2018kdj}, we develop a path-integral optimization formulation for the computation complexity of warped CFTs.

\subsection{RG flow with two fields, the case of WCFTs}\label{sec:twocutoffs}

To understand the mechanism of AdS/CFT and the emergence of spacetimes out of information, the idea of multi-scale renormalization ansatz (MERA) has been applied in \cite{Swingle:2009bg}.  To further study such idea, specially for the continuum limit, the optimization procedure of Euclidean path-integral which can evaluate a CFT wave functional has been introduced in \cite{Caputa:2017yrh,Caputa:2017urj}. This optimization would be performed by minimizing the Liouville action.

We could then apply the same mechanism for the case of a chiral, warped $\text{CFT}_2$. Our conjecture is to replace the Liouville action with the chiral Liouville action introduced in previous section, and then check how different symmetries, anomalies and chirality would affect the process of emergence of spacetime out of boundary field theory.  For that, one needs to understand the microscopic structure of complexity for the case of warped CFTs where the left and right modes behave differently. So we would like to understand how a warped CFT state $\ket{\Psi_\Sigma} $, and then the warped geometry, could be constructed using the tensor network description and from a similar optimization procedure.

A relation similar to \cite{Takayanagi:2018pml} in the form of 
 \begin{gather}
 e^{C(M_\Sigma) } \ . \ \Psi_\Sigma [\varphi_0(t^+)] = \int \left[ \prod_{r \in M_{ \Sigma}}  D\varphi (r)\right]
  e^{-S_{M_\Sigma}^{\text{WCFT}} [\varphi ]}  \prod_{x \in \Sigma } \delta (\varphi (t^+) - \varphi_0 (t^+) ), 
  \end{gather}
for the case of warped CFT states should be derived . In the above relation, $S_{M_\Sigma}^{\text{WCFT} }[\varphi] $ should be replaced by the action for WCFT theories, which would be the chiral Liouville action as discussed previously or other Lagrangians for warped CFTs which we will introduce later. The optimization procedure then would be performed by minimizing the functional $C_L(M_\Sigma)$ with respect to the two fields of $\rho$ and $h$.

 \begin{figure}[ht!]
\centering
\includegraphics[width=0.25\textwidth]{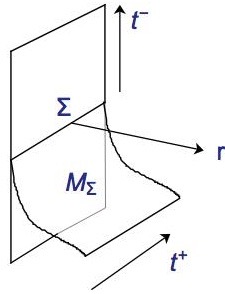} \hspace{1cm}
\includegraphics[width=0.25\textwidth]{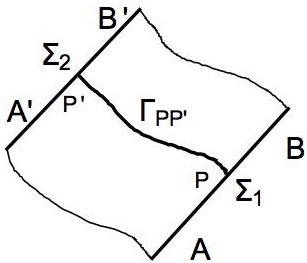}
\caption{ The relation between the codimension one surface $M_\Sigma$  and where it ends on $\Sigma$ could be the same as those of \cite{Takayanagi:2018pml} shown in the left side and the quantum circuit modeled by the path-integration on $M_{\Sigma_1 \Sigma_2 }$ connecting $\Sigma_1$ to $\Sigma_2$ is modeled in the right side. We show however later that a twist is needed to be considered for the WCFT/WAdS case.}\label{fig:picmanifold2}
\end{figure}

The plot of the manifold and its boundary is shown in figure \ref{fig:picmanifold2}. Note that in the WCFT case we don't have the complete conformal invariance, therefore, the state $\ket{\Psi_\Sigma}$ after path integration could depend not only on $\Sigma$ but the whole $M_\Sigma$. However, in the warped $\text{CFT}_2$ case, there is still a warped analogue of the conformal transformation connecting the metrics of the two surfaces $M_\Sigma^{(1)}$ and $M_\Sigma^{(2)}$ which will be used to simplify the process.

So first, deriving the corresponding equation of motion of $(2.10)$ in \cite{Caputa:2017yrh}, for the case of warped CFT and chiral Liouville action would lead to
\begin{gather}\label{eq:motionEQ}
\partial_+ \partial_- \rho+\frac{3k}{2c}h+h (\partial_- \rho)^2+\frac{\Lambda}{8} e^{2\rho}=0,
\end{gather} 
where we also have $\partial_- h=0$. One could see from this equation how these two fields are coupled, behave differently in left versus right, and how central charge and Kac-Moody level simultaneously play a role. After choosing a boundary condition, one needs to solve these coupled equations for the field $h$ and $\rho$ which then lead to the time-slice of warped $\text{AdS}_3$.

Then, the renormalization group flow for the case of WCFTs could be considered in order to understand further its path-integral optimization complexity. In fact, for this case, there should be two position dependent cut offs which would be the functions of coordinates $(t^+, t^-)$ as
\begin{gather}
\Lambda_1=e^{\rho(t^+,t^-)}, \ \ \ \  \ \ \ \ \ \Lambda_2=e^{h(t^+,t^-)}.
\end{gather}
If we want to have a theory that would not change after optimization, both cutoffs should satisfy the RG equation,
\begin{gather}
\Lambda_i \frac{d\lambda_i(\Lambda_i)}{d \Lambda_i}=\beta (\lambda_i(\Lambda_i) ),
\end{gather}
where $\beta$ is the beta function.

The Callan-Symmanzik equation for the partition function $Z$ and also the correlation function would satisfy the following equations
\begin{gather}
\left( \Lambda_\rho \frac{d}{d\Lambda_\rho} +\beta(\lambda_\rho)\frac{d}{d\lambda_\rho} +\xi (\lambda_\rho) \right) Z[\Lambda_\rho, \lambda_\rho]=0,\nonumber\\
\left( \Lambda_h \frac{d}{d\Lambda_h} +\beta(\lambda_h)\frac{d}{d\lambda_h} +\xi (\lambda_h) \right) Z[\Lambda_h, \lambda_h]=0.
\end{gather}

The $\xi(\lambda)$ functions here would break the conformal symmetry leading to the WCFT case.

We then need to understand how to couple these two RG equations for these two fields. As for a general partition function $Z$ for the whole system, we could have the following relation
\begin{gather}
\left( \Lambda_\rho \frac{d}{d\Lambda_\rho} +\Lambda_h \frac{d}{d\Lambda_h}+\beta(\lambda_\rho)\frac{d}{d\lambda_\rho}+\beta(\lambda_h)\frac{d}{d\lambda_h} +\xi (\lambda_\rho)+\xi (\lambda_h) \right) Z[\Lambda_\rho, \Lambda_h,\lambda_\rho, \lambda_h]=0.
\end{gather}

In \cite{Aharony:2018mjm, Narovlansky:2018muj}, the renormalization group flow equations for field theories with quench disorder as an example of non-relativistic, coupled RG flow have been studied where their couplings could vary randomly in space. In the case they have studied, the couplings were also mixed together, similar to the situation for WCFTs. So their methods could be applied here as well. So in the next parts, using various methods, we try to understand these coupled renormalization group flows for the case of WCFTs better.

\subsection{Local RG Flow in the locally chiral gauge}

As mentioned, Liouville action is always written in the conformal gauge, it is local and has a left and right Virasoro algebra. However,  in the chiral gauge, the chiral Liuoville theory has only the right-moving Virasoro Kac-Moody algebra and it is in fact semi-local.

Also, for the case of CFT we have the ground state and its Virasoro descendants. For the case of warped CFTs, however, we just have a left Virasoro and a right $U(1)$ Kac-Moody algebra. As the right and left descendants behave differently,  the Euclidean path integral optimization process that prepares the left and right moving states would be different.

For the ground state of CFT we have the relation
\begin{gather}
\Psi (\tilde{\varphi} (x) ) =\int e^{-S_{\text{CFT} (\varphi) }} \prod_{x}  \prod_{\delta <z< \infty} D\varphi(z,x) \ \Big |_{\varphi(\delta,x)= \tilde{\varphi}(x) },
\end{gather}
which computes the weight of a field configuration $\tilde{\varphi} (x)$. The Virasoro descendants could be computed by changing the cutoff surface to $z=\delta$.  For the chiral Liouville case, there would be two parameters, one free continuos parameter $k$ which determines the central charge and the other one is the discrete parameter $\xi$, which determines the sign of $U(1)$ Kac-Moody level.

To understand the setup better, we consider the metric of the field theory in the locally chiral gauge as \cite{Compere:2013aya}
\begin{gather}\label{eq:WCFT1}
ds^2=-e^{2\rho(t^+,t^-)} \left( dt^+ dt^--h(t^+,t^-) (dt^+)^2  \right),
\end{gather}
where $\partial_- h=0$. Note that in this gauge, the coordinates are not orthogonal. Also, here the field $\rho(t^+, t^-)$ couples the left and right coordinates $dt^+$ and $dt^-$.

As this metric is symmetric, of course at a single point one could diagonalize it in the following way
\[
   ds^2= \frac{e^{2\rho}}{2}
  \left[ {\begin{array}{cc}
   h+\sqrt{1+h^2} & 0 \\
   0 & h-\sqrt{1+h^2} \\
  \end{array} } \right].
\]

In this new coordinate $(t'^+,t'^-)$, the size of a lattice could locally be defined as
\begin{gather}
\Delta l_1=e^{-\rho} \sqrt{\frac{2}{h+\sqrt{1+h^2} } }, \ \ \ \  \ \ \ \ \ \Delta l_2=e^{-\rho} \sqrt{\frac{2}{h-\sqrt{1+h^2} } },
\end{gather}
where $\rho (t'^+,t'^-)$ and $h(t'^+,t'^-)$ are local functions of $t'^+$ and $t'^-$.

Therefore, unlike the case of \cite{Bhattacharyya:2018wym}, in the Wilsonian RG flows here, we should have two cut off scales $\Lambda_\rho$ and $\Lambda_h$ as
\begin{gather}
\Lambda_{\rho_0}= e^{\rho_0}, \ \ \ \ \ \ \ \  \ \Lambda_{h_0}=e^{\rho_0} \sqrt{h_0}.
\end{gather}

In order not to change the theory after optimization, we should then have the following two conditions,
\begin{gather}
\frac{\lambda_\rho}{\delta \rho}=\beta ( \lambda_\rho),\ \ \ \ \ \ \ \ 
\frac{\lambda_h}{\delta h }=\beta(\lambda_h),
\end{gather}
where the two beta functions are
\begin{gather}
\beta( \lambda_\rho)= (\Delta_\rho-2) \lambda_\rho+O(\lambda_\rho^2), \ \ \ \ \ \ \ \ \
\beta(\lambda_h)= (\Delta_h-2) \lambda_h+O(\lambda_h^2).
\end{gather}

For any operator with the dimension $\Delta$, the beta function could be written as \cite{Bhattacharyya:2018wym},
\begin{gather}
\beta (\lambda)=(\Delta-2)\lambda+O(\lambda^2).
\end{gather}

Now in order to study the RG flow locally, we make the cut offs position dependent, so we have 
\begin{gather}
\frac{\lambda_\rho(t^+,t^-)}{\delta \rho (t^+,t^-)}=\beta[\lambda_\rho]^{(t^+, t^-)}_{(t^+,t^-)},\ \ \ \ \ \ \
\frac{\lambda_h(t^+,t^-)}{\delta h (t^+,t^-)}=\beta[\lambda_h]^{(t^+, t^-)}_{(t^+,t^-)},
\end{gather}
where
\begin{gather}
\beta[\lambda_\rho]_{(t^+,t^-)}^{(t^+,t^-)}= (\Delta_\rho-2) \lambda_\rho(t^+,t^-) \ . \ \delta(t^+-t'^+)\delta(t^--t'^-)+O(\lambda_\rho^2), \nonumber\\ 
\beta[\lambda_h]_{(t^+,t^-)}^{(t^+,t^-)}= (\Delta_h-2) \lambda_h (t^+,t^-) \ . \ \delta(t^+-t'^+)\delta(t^--t'^-)+O(\lambda_h^2).
\end{gather}

The higher order terms are non-local. 

As the partition function for the warped CFT depends on all the parameters $\phi, \lambda_\phi, h, \lambda_h$, the Callan-Symanzik equation locally could be written as
\begin{align}
&\bigg (\frac{\delta}{\delta \rho (t^+, t^-)} +\int dt'^+ dt'^- \beta[\lambda_\rho]^{(t'^+,t'^-)}_{(t^+,t^-)} \ . \  \frac{\delta}{\delta \lambda_\rho (t'^+,t'^-) }+\xi[\lambda_\rho]_{(t^+,t^-)}+ \nonumber\\ &
\frac{\delta}{\delta h (t^+, t^-)} +\int dt'^+ dt'^- \beta[\lambda_h]^{(t'^+,t'^-)}_{(t^+,t^-)} \ . \  \frac{\delta}{\delta \lambda_h (t'^+,t'^-) }+\xi[\lambda_h]_{(t^+,t^-)} + \nonumber\\ &\text{coupling term $f(\lambda_\rho) g(\lambda_h)$} \bigg) Z[\rho, \lambda_\rho, h, \lambda_h]=0.
\end{align}

We propose that for the warped CFT case, the normalization functional $N[e^{2\rho},\lambda_\rho, h, \lambda_h]$ in the relation
\begin{gather}
\Psi_{g=e^{2\rho}, \lambda_\rho, h ,\lambda_h} [\varphi(\tau)] =e^{N[e^{2\rho},\lambda_\rho,h,\lambda_h] -N[e^{2\rho_0}, \lambda_{\rho_0},h_0, \lambda_{h_0} ]} \ . \ \Psi_{g=e^{2\phi_0} ,\lambda_0 [\varphi (x)]  },
\end{gather}
would be a function of the chiral Liouville action \cite{Compere:2013aya},

\begin{gather}\label{eq:chiralliouville}
N[e^{2\rho},\lambda_\rho,h,\lambda_h]=S_{CL}[\phi]= \frac{c}{12\pi} \int d^2 x \left(\partial_+ \partial_- \rho -\frac{\lambda_\rho}{8} e^{2\rho} +h(\partial_- \rho)^2+ [\partial_- h \partial_- \rho] -\frac{6}{c} h \Delta(\lambda_h) \right).
\end{gather}

In relation \ref{eq:chiralliouville}, the fourth term would not be important, since in the chiral gauge we have $\partial_- h=0$. Also, the parameter $\Delta/2$ is the left-moving energy density and as could be seen from equation (3.8) in \cite{Compere:2013aya}, it is related to the coupling constant $\lambda_h$. In the metric of $\text{AdS}_3$, it would actually be related to $\alpha$ and from the algebraic point of view, to the Kac-Moody level $k$.

To understand the significance of each term for the tensor network, and how they would be related to the number of isometries and unitaries, or any new necessary gates such as chiraloens, we review the Hamiltonian of the chiral Liouville theory.

Taking the canonical momenta as
\begin{equation}\label{eq:WCFTmomenta}
\begin{split}
\frac{12\pi}{c} \frac{\delta S_{cL}}{\delta \partial_- \rho} \equiv \Pi_\rho &=(\partial_+ \rho+2  \partial_- \rho+\partial_- h), \nonumber\\
\frac{12\pi}{c} \frac{\delta S_{cL}}{\delta \partial_- h} \equiv \Pi_h &=\partial_- \rho,
\end{split}
\end{equation}
the Hamiltonian could be written as \cite{Compere:2013aya}.
\begin{gather}
H=\int dt^+ \left(\Pi_h \Pi_\rho- \Pi_h \partial_+ \rho -h {\Pi_h}^2 +\frac{\Lambda}{8} e^{2\rho} +\frac{\Delta}{k} h \right).
\end{gather}

From this form, one could then see that the first three terms are related to the kinematics and therefore they are related to the number of isometries in the warped conformal field theories \cite{Czech:2017ryf}. The fourth term is the potential term and therefore, it is related to the number of unitaries, and also to the number of tensors corresponding to the field $\rho$ \cite{Caputa:2017yrh} in the bulk. The last term which is the only one related to the Kac-Moody level $k$ would be related to the number of unitaries for the field $h$. Therefore, the number of the new gates would be counted by the last term. So from this Hamiltonian form, one could note that the middle term in \ref{eq:chiralliouville}, which is a deforming term, plays the role of isometries as it is related to the kinetic term for the field $h$. 

Another interesting point worths to mention, is that in the usual case of AdS, the time coordinate in a gravity dual emerges from the complexity of the Hamiltonian circuit in the dual field theory \cite{Takayanagi:2018pml}. For the warped AdS case, however, as one could see from the form of Hamiltonian, the circuits acting on the right moving and left moving modes are very different. As a matter of fact, the number of gates acting on the right moving relative to left moving is a function of the parameters $h$ and $\Delta$.

Understanding the significance of each term of the chiral Liouville gravity with respect to its tensor network structure and using the conjugate momenta \ref{eq:WCFTmomenta} and chiral Hamiltonian, we could then even propose the form of ``entanglers"  for WCFTs. 

This functional only has been introduced for a free scalar field in one spatial dimension in a $2d$ CFT. Now we propose the following entangler functional $K(s)$ for the $2d$ WCFTs, as
\begin{gather}
K(s)=-\frac{i}{2} \int dt^+ dt^- g(t^+-t^-,s) \Big [ \Psi_h (t^+) \Psi_\rho(t^-) - h \Psi_h(t^+) \Psi_h(t^-)
-\partial_+ \rho(t^+) \Psi_h (t^-) \Big]. 
\end{gather}

In this relation, similar to \cite{Franco-Rubio:2019nne}, the $\Psi$s are the annihilation operator. Note also the cut offs for each field $\rho$ and $h$ are actually different. So we have
\begin{gather}
\Psi_\rho (t^i) \equiv \sqrt{\frac{\Lambda_\rho}{2}} \rho(t^i)+i \sqrt{\frac{1}{2\Lambda_\rho}} \Pi_\rho (t^i),\ \ \ \ 
\Psi_h (t^i) \equiv \sqrt{\frac{\Lambda_h}{2}} h(t^i)+i \sqrt{\frac{1}{2\Lambda_h}} \Pi_h (t^i),\nonumber\\
\end{gather}
and $g (t,s)$ is a profile at scale $s$ where a suitable form should be considered for the states of WCFTs.  However, as mentioned in \cite{Franco-Rubio:2019nne}, there is a lot of freedom in choosing the entangler $K$ as many other forms for this operator and also various profiles for $g$ could be proposed. We leave determining the exact class of these profiles and entangler functions for the case of WFTs to future works.

\subsection{Actions for Warped CFTs}\label{sec:ActionWCFT}
In \cite{Hofman:2014loa,Jensen:2017tnb}, the Lagrangians for the WCFT theories have been constructed which could help to get a better picture of the path integral optimization complexity for the warped CFTs. These theories are explained as follows.

\subsubsection{Scalar Case}

For a free scalar field $\varphi$, in \cite{Jensen:2017tnb}, the Lagrangian has been constructed as
\begin{gather}
S=\int dt^+ dt^- \sqrt{\gamma} \Big \{ \frac{1}{2} (\omega^\mu \partial_\mu \varphi )^2+\frac{1}{4}  \Big( \omega^\mu \partial_\mu N -\frac{N^2}{2} \Big ) \varphi^2 -\frac{m^2}{2} \varphi^2 \Big \},
\end{gather}
where $N=\epsilon^{\mu \nu} \partial_\mu n_\nu$. The first term is the kinetic term, the last term is the potential term and the middle one models the chiral coupling to the curvature. In the flat space this action would become \cite{Jensen:2017tnb}
\begin{gather}
S=\int  dt^+ dt^- \Big \{ \frac{1}{2} (\partial_+ \varphi)^2 -\frac{m^2}{2} \varphi^2 \Big \}.
\end{gather}

In the above action the kinetic term which corresponds to the number of disentanglers or isometries in the tensor network structure has only a right moving direction. Therefore, the disentanglers only would influence these right-moving modes leading to such a mathematical relation. 

The mass here could be considered as a small, perturbative parameter which is a function of both of the coordinates, in the form of $m(t^+,t^-)$. Then, similar procedure to \cite{Bhattacharyya:2018wym} could be applied to derive the normalization functional and the path integral complexity for this theory. Note here though the operator $\varphi^2$ is not a primary operator.

The solution for the the scalar field of this action is
\begin{gather}
\varphi= \varphi_+(t^-) e^{im (t^+,t^-) t^+} +\varphi_- (t^-) e^{- i m(t^+,t^-) t^+}, 
\end{gather}
where the parameter $m$ is setting the Kac-Moody level. Also, these theories admit an infinite number of ``exactly marginal \textit{non-local} deformations". 

The stress tensor $T$ and momentum $P$, as the most important physical parameters of the theory, have been calculated in \cite{Jensen:2017tnb}, leading to the following relations,
\begin{flalign}
T &=\frac{1}{4\pi} ( \partial_- \varphi \partial_+ \varphi- \varphi \partial_- \partial_+ \varphi), \nonumber\\
P &= \frac{1}{4\pi} ((\partial_+ \varphi)^2 -m^2 \varphi^2).
\end{flalign}

To search for the normalization functional in the path integral complexity of warped CFTs, one then should use the correlation functions, or specifically the retarded Green's function for the primary operators of WCFTs calculated in \cite{Song:2017czq}, as
\begin{gather}
G_R(\omega) \sim \beta^{1-2\Delta} \frac{ e^{\frac{ Q\bar{q} \beta  }{2} } } {\sin \left (\pi \Delta-\frac{i Q \bar{q} \beta}{2 }
  \right )} \frac{\Gamma \left ( \Delta+i \frac{\omega+Q \bar{q}}{2\pi/\beta} \right) }{\Gamma \left ( 1-\Delta+i \frac{\omega+Q \bar{q}}{2\pi/\beta} \right) },
\end{gather}
where $\Delta$ is the weight and $Q$ is the charge of the primary operator which defines the state $\mathcal{O} (0,0) \sim \ket{\Delta,Q}$. Also, the two-point function for the twist operator $\Phi(X,Y)$ which has dimension $\Delta_n$ and charge $Q_n$ has been found in \cite{Song:2017czq} as
 \begin{gather}
 \langle \Phi_n (X_1, Y_1) \Phi^\dagger_ {n} (X_2,Y_2) \rangle_\mathcal{C} \sim e^{i Q_n \left(Y_1-Y_2+\frac{\bar{\beta}-\alpha}{\beta} (X_1-X_2) \right)} \left(\frac{\beta}{\pi} \sinh \frac{\pi (X_1-X_2) }{\beta} \right)^{-2 \Delta_n} ,
 \end{gather}
where here $Y$ specifies the classically $U(1)$ preferred axis, and $X$ denotes the quantum anomaly selected axis with a scaling $SL(2,R)$ symmetry. The thermal identification here is $(X,Y) \sim (X+i \beta,Y-i\bar{\beta})$. Also, $\Delta_n$ and $Q_n$ could be written as
\begin{gather}
\Delta_n= n  \left( \frac{c}{24}+\frac{\mathcal{L}_0^{vac} }{n^2} +\frac{i \mathcal{P}_0^{vac} \alpha }{2n  \pi}-\frac{\alpha^2 k}{16 \pi^2} \right), \ \ \ \ Q_n=n \left(-\frac{\mathcal{P}_0^{vac} }{n}-i \frac{k \alpha}{4\pi} \right ).
\end{gather}

Knowing the Green's function, similar to equation $(A.10)$ of \cite{Bhattacharyya:2018wym}, the normalization functional for calculating the path-integral optimization could be found as
\begin{gather}\label{eq:CLM}
N \sim S_{CL}-\frac{m^2}{2} \int dt^+ dt^- \sqrt{g} G_R(\omega).
\end{gather}

From this, one could see how each parameter such as the Kac-Moody parameter, central charge, the tilt angle $\alpha$, or the mass term change the path integral complexity. 

Additionally, since the correlation functions of warped CFTs have already been found in \cite{Song:2017czq}, then using similar relations, $(5.14)$ and $(5.15)$ of \cite{Bhattacharyya:2018wym}, for the WCFTs, one could find the first and second order of the normalization functional in terms of the correlators of WCFTs on a half plane with the boundary at $t^+=\epsilon$ as 
\begin{gather}
N \sim S_{CL} +N_{1pt}+N_{2pt}+....
\end{gather}
where
\begin{gather}
N_{1pt} \sim   \int dt^+ dt^- e^{2 \rho (t^+,t^-)} \lambda_\rho (t^+) \lambda_h \langle O \rangle.
\end{gather}

For the second order, the following relation could be written in terms of the two point correlation function, as
\begin{flalign}
N_{2pt} & \sim \left( \prod_{i=1}^2 \int dt^+_i dt^-_i \sqrt{g} \right) \lambda_\rho (t^+_1) \lambda_\rho (t^+_2) \lambda_h (t^+_1) \lambda_h (t^+_2)\nonumber\\&
\times \Big( \langle O(t_1^+, t_1^-) O(t_2^+, t_2^-) \rangle- \langle O(t_1^+,t_1^-) \rangle \langle O(t_2^+,t_2^-) \rangle  \Big).
\end{flalign}

The exact normalization functional which has a more complicated form would depend on both fields $\phi$ and $h$, the coupling parameter for each of them, and also the Kac-Moody parameter, as well as central charge which could be calculated from the relation \ref{eq:CLM}. In this process, further methods of \cite{Pal:2017ntk}, for calculating the heat kernel of such Lorentz-violating theories would be very helpful.

\subsubsection{The Hofman/Rollier theories}
Similar to \cite{Bhattacharyya:2018wym}, other forms of actions could be considered for the warped CFT case. For the first time, in \cite{Hofman:2014loa}, using warped geometry, Hoffman and Rollier wrote two Lagrangians for the WCFTs, which were a theory of ``Weyl fermions" with a Lorentz-violating mass and also a warped ``bc" system which then for the case of free scalar WCFTs were reviewed and extended further in \cite{Jensen:2017tnb}.  For each of these theories, similar to \cite{Bhattacharyya:2018wym}, the optimization procedures and then, the derivation of the wave functional behaviour could be studied. 

Considering an anti-commutating, complex spinor field $\Psi=(\Psi^-, \Psi^+)$, with its conjugate $\bar{\Psi}$ one could write the action in the following form, where the mass is a function of the coordinates as in \cite{Bhattacharyya:2018wym} 

\begin{gather}\label{eq:warpedWeyl}
S_{\textit{weyl}}=\int dt^+ dt^- \sqrt{\gamma} \{ i \overline{\Psi}^- \omega^\mu \partial_\mu \Psi^-+m(t^+,t^-) \overline \Psi^- \Psi^-\}.
\end{gather}

In the above action, $\Psi^-, \Psi^+$ are the two components of an anti-commuting field $\Psi$ which under boosts transform as
\begin{gather}
\bar{\Psi}^+ \to \bar{\Psi}^+ + \frac{\psi}{2} \bar{\Psi}^-, \ \ \ \ \ \bar{\Psi}^- \to \bar{\Psi}^-,
\end{gather}
and also
\begin{gather}
\Psi^- \to e^{-\frac{\Omega}{2}} \Psi^-, \ \ \ \ \ \bar{\Psi}^- \to e^{-\frac{\Omega}{2}} \bar{\Psi}^-.
\end{gather}

The dimension of the fermion field here is $1/2$.

It worths to refer here to another action called warped $bc$ theory which has a real spinor $\Psi$  written in the form
\begin{gather}\label{eq:warpedBC}
S_{bc}= \int d^2x \sqrt{\gamma} \{ i \Psi^- \nu^\mu \partial_\mu \Psi^--2i \Psi^+ \omega^\mu \partial_\mu \Psi^- -2m \Psi^+ \Psi^- \}.
\end{gather}

In \cite{Castro:2015uaa}, the partition function for a free warped Weyl fermion with a complex anti-commuting field $\Psi$ and with the action
  \begin{gather}
 I= \int dt d\varphi (i \bar{\Psi} \partial_+ \Psi+ m \bar{\Psi} \Psi ),
 \end{gather}
 has been calculated as
 \begin{gather}
 \hat{Z}(z \big | \tau) =Tr \left( e^{2 \pi i z \hat {P}_0} e^{-2 \pi i \tau (\hat {L}_0 -\frac{c}{24} ) }  \right),
 \end{gather}
 where $\hat{L}_0$ and $\hat{P}_0$ are the plane generators. For this system, as the first approximation, the functional  which should be minimized could be given by the log of this partition function, $S \propto \log  \hat{Z} $, which would depend on both of the generators of the algebra.
 
In fact in \cite{Hofman:2014loa}, it has been shown that the partition function of WCFTs defined on a non-trivial background, up to a quantum anomaly, has a notion of warped Weyl symmetry as
\begin{gather}
Z[A_\mu, \bar{A}_\mu] \sim Z[(1+\gamma) A_\mu, \bar{A}_\mu +\nu A_\mu],
\end{gather}
where $\gamma$ and $\nu$ are arbitrary deformation parameters which only depend on the space-time coordinates.

There would also be another interesting example of dual bulk geometry for the warped CFTs, which is a $SL(2) \times U(1)$ Chern-Simons model, dubbed  \textit{lower spin gravity}, written in the following form \cite{Hofman:2014loa,Azeyanagi:2018har}
\begin{gather}\label{eq:lowerSpinGravity}
S=\frac{k}{2} \int_{bulk} B^1 \wedge dB^1+B^2 \wedge dB^2-B^3 \wedge dB^3+2 B^1\wedge B^2 \wedge B^3- \xi \int_{bulk} \bar{B} \wedge d\bar{B}.
\end{gather}

In the above action, we have $k=\frac{2\kappa}{c}$, where $\kappa$ is an overall normalization and $\xi=+1, 0 , -1$ is the sign of the last term which depends on the parameter of the theory.  So this bulk theory similar to the dual WCFT case, has a free continuous parameter $k$ which is dual to the central charge and a discrete parameter $\xi$ which determines the level of $U(1)$ Kac-Moody algebra. Therefore, this action could be considered as the minimal bulk gravity theory dual to WCFTs. The parameters of this action could be then connected to the parameters of warped $\text{AdS}_3$ noted in \cite{Hofman:2014loa}.

The main point here is that, as mentioned in \cite{Jensen:2017tnb},  both of the actions \ref{eq:warpedWeyl} and \ref{eq:warpedBC} could admit an infinite number of exactly marginal deformation operators and it means both are on the ``brink" of being non-local along $x^+$, so building an RG flow for these specific theories using a ``semi-local" entangler $K$ would be rather challenging. However, still some constraints on WCFTs could be considered which tune away such non-localities and therefore still building a physical tensor network would be possible.

As the Hofman/Rollier theories describe a fermion system and our objective is to calculate the discretized path integral optimization and finding the tensor network structure for such systems, several subtleties involving issues about fermions on a lattice should be considered. First, as has been shown in \cite{RandjbarDaemi:1995cq,Kaplan:1992bt}, the chiral theories are renormalizable both within and outside of perturbation theory which lead to the conclusion that the theory in fact would be physical.

Another issue is the fermion doubling problem \cite{2004PrPNP..53..373C} which arrises when one tries to put fermionic fields on a lattice leading to the appearance of spurious states such as $2^d$ fermion particles for each previous fermion, where here $d$ is the number of discretized dimension. There is also the Nielsen-Ninomiya theorem \cite{NIELSEN1981219}, which shows that a local, real free fermion lattice action which has chiral and translational invariance would necessarily has fermion doubling. However, there are various ways to circumvent this problem, such as ``perfect lattice fermions",  ``Wilson fermions", ``twisted mass fermions", ``domain wall fermions", ``overlap fermions", ``interacting fermions", "\textit{staggered fermions}", etc.  Specifically the staggered fermion introduced by Susskind and Kogut \cite{Kogut:1974ag}, could be of interest as it introduces a novel "nonlocal action" similar to the actions presented here.

Also, note that for the construction of tensor networks, one could use additional tensors such as ``\textit{smoothers}" introduced in \cite{Milsted:2018vop}, or ``\textit{chiraleon}" introduced here which could help to circumvent some of these problems. 

In \cite{PhysRevB.95.245127}, already the tensor network for fermionic topological quantum states has been worked out where a Grassmann number tensor network ansatz  for a fermionic twisted quantum model has been formulated. There, the fermionic projected entangled pair state fPEPS for a simple string-net model has been shown. One could also extend their procedure for the above chiral system.

\section{Derivation of actions from geometries}\label{sec:DerAction}

In the previous section, we generally tried to start from the boundary warped CFTs and using tools such as path-integral optimization complexity, the possible structure of RG flow and the form of Hamiltonian and different Lagrangian of the theory, understand the emergence of warped spacetimes out of a chiral theory. In this section, however, we start from the other side of the story. We will first take the metric of warped $\text{AdS}_3$ and by using the form of the induced metric on a boundary profile derive a deformed chiral action.

First, we review the procedure of deriving Liouville action from $\text{AdS}_3$ similar to the approach of \cite{Takayanagi:2018pml} and then we repeat it for the warped AdS and chiral Liouville action.

\subsection{Deriving Liouville action from $\text{AdS}_3$ geometry}\label{sec:DesAdS}
Considering the Euclidean Poincare $\text{AdS}_3$ as
\begin{gather}
ds^2=R^2_{\text{AdS}} (dz^2+dT^2+dX^2)/z^2,
\end{gather}
we could take the boundary on the cut off point at $z \ge \epsilon . e^{-\tilde{\phi} (T,X) }$. We also have the relation
\begin{gather}
 \frac{R^2}{z^2}= \frac{e^{2\tilde{\phi} } }{\epsilon^2}.
\end{gather}

The induced metric on the boundary $z=\epsilon \ . \ e^{-\tilde{\phi} } $ would be
\begin{gather}\label{eq:inducedAdS}
ds^2=\frac{e^{2\tilde{\phi} } }{\epsilon^2} \left [ \left( 1+\epsilon^2 e^{-2 \tilde{\phi} } (\partial_T \tilde{\phi} )^2 \right) dT^2 +2 \epsilon^2 e^{-2 \tilde{\phi} } (\partial_T \tilde {\phi} )(\partial_X \tilde {\phi} ) dT dX +\left(1+\epsilon^2 e^{-2 \tilde{\phi} } (\partial_X \tilde{\phi})^2 \right) dX^2 \right], 
\end{gather}
and then the extrinsic curvature on this boundary could be found as
\begin{gather}\label{AdSextrinsic}
K= R_{\text{AdS}}^{-1} \  .  \left(  2- \epsilon^2 e^{-2 \tilde{\phi} }(\partial_T^2+\partial_X^2) \tilde{\phi}    \right).
\end{gather}

Then, the bulk gravity action could be calculated as
\begin{gather}\label{actionwhole}
I_G=\frac{1}{4\pi G_N R^2_{\text{AdS}}} \int_N \sqrt{g} -\frac{1}{8\pi G_N} \int_M \sqrt{\gamma} K\nonumber\\ = -\frac {c}{12 \pi} \int dT dX \left[ \frac{e^{2\tilde {\phi} }}{\epsilon ^2} +(\partial_T \tilde{\phi} )^2 +(\partial_X \tilde{\phi} )^2 \right ].
\end{gather}

The metric of CFT on the boundary would be
\begin{gather}
{ds^2}_{\text{CFT}}= e^{2\phi(t,x)} (dt^2+dx^2).
\end{gather}

To go from the coordinates $(T,X)$ to $(t,x)$ we need to calculate the Jacobian $J$ such that $dT dX= J dt dx$. This $J$ could be found as
\begin{gather}\label{eq:jacobian}
e^{2\phi} = J \ \frac{e^{2 \tilde{\phi} } } {\epsilon^2} \ \sqrt{1+\epsilon^2 e^{-2 \tilde {\phi}} \left( (\partial_T \tilde{\phi} )^2+(\partial_X \tilde{\phi} )^2  \right) }.
\end{gather}

Till here we actually derived a relation which is very similar to the equation (22) of \cite{Camargo:2019isp}
\begin{gather}
\mathcal{D}_L \simeq \int dt dy \frac{1}{\epsilon^2} \sqrt{a^2+ \epsilon^2 \eta _{(\partial a)^2} (\partial_y a)^2+\epsilon^2 \eta_{(\partial b)^2} (\partial_y b)^2+...   },
\end{gather}
where $a$ and $b$ are the parameters of the metric
\begin{gather}
ds^2=\left( a(t,y)^2+b(t,y)^2 \right )dt^2+2b(t,y) dt dy+dy^2,
\end{gather}
where the constant Euclidean time $t$ slices are flat lines.

In fact, the connection with the DBI action 
\begin{gather}
\mathcal{D}_{DBI} \simeq \int d^2 \chi \sqrt{\text{det} (g_{\mu \nu}+\epsilon^2 \partial_\mu \chi \partial_\nu \chi ) }, 
\end{gather}

for the case of zero angle $\sigma$ where $t'=-\dot{t} \tan \sigma$ has been proposed in \cite{Camargo:2019isp}. 
The dual action of DBI case for the warped case could be considered.

So getting back to our calculations, using the Jacobian \ref{eq:jacobian}, one could write the first term of $I_G$ as 
\begin{gather}
\int dT dX \frac{e ^{2 \tilde{\phi} } }{\epsilon^2} \simeq \int dt dx \left[ e^{2\phi} -\frac{1}{2} \left( (\partial_t \phi)^2 +(\partial_x \phi)^2 \right)  \right ].
\end{gather}

Putting this into \eqref{actionwhole} would lead to
\begin{gather}
I_G^E=-\frac{c}{24\pi} \int dt dx \Big [(\partial_t \phi )^2+(\partial_x \phi)^2+2e^{2\phi} \Big],
\end{gather} 
as it was shown in the appendix section of \cite{Takayanagi:2018pml}. Now we could repeat these calculations for the warped $\text{AdS}_3$.

Before doing that, we should remind that, from the perspective of \cite{Camargo:2019isp}, the Liouville action is a particular cost function. The chiral Liouville action then could be considered as a particular cost function for the non-local and Lorentz-breaking theory of warped CFTs. By fine-tuning the cost function in the case of Liouville, one could recover a complexity measure of the same form. Now, we could check how this fine tuning would work for the case of chiral Liouville action.

\subsection{Deriving chiral Liouville action from warped $\text{AdS}_3$}\label{sec:ChiralDerAct}
In this section, we repeat the previous calculations for the case of various warped $\text{AdS}_3$ to check what kind of geometries could be derived and therefore to get further information about the kind of quantum circuits which could be built for a chiral, warped geometry.

The metric of warped $\text{AdS}_3$ in the Poincare coordinate could be written as 
\begin{equation}
\begin{split}
ds^2 &=L_{WAdS}^2 \left ( \frac{dr^2}{r^2}-r^2 {dX^-}^2 +\alpha^2 (dX^+ + r dX^-) ^2 \right) \nonumber\\
&=L_{WAdS}^2 \left( \frac{dr^2}{r^2} + {dX^{-}}^2  \left( \alpha^2 r^2-r^2 \right) +\alpha^2 {dX^+}^2 +2 \alpha^2 r dX^+ dX^-  \right),
\end{split}
\end{equation}
where comparing with the notation of \cite{Anninos:2008fx}, one could write
\begin{gather}
L_{WAdS}=\frac{\ell^2}{\nu^2+3}, \ \ \ \ \ \ \alpha^2=\frac{4\nu^2}{\nu^2+3}.
\end{gather} 

Again, considering a simple and naive cutoff surface as $r= \epsilon \ . \ e^{- \tilde{\phi} (X^+, X^-)  } $, one could find the induced metric on this surface as
\begin{gather}
\begin{split}
ds^2 &= \frac{e^{2\tilde{\phi} }}{\epsilon^2} L^2_{WAdS} \Big [  \Big (r^2 \alpha^2+\epsilon^2 e^{-2\tilde{\phi} }  (\partial_{X^+} \tilde{\phi})^2    \Big) {dX^+}^2+2\epsilon^2 e^{-2 \tilde{\phi} }\ \partial_{X^+} \tilde{\phi} \ \partial_{X^-} \tilde{\phi} \ dX^+ dX^-  \nonumber\\& + \Big( r^4 (\alpha^2-1) +\epsilon^2 e^{-2\tilde {\phi}} (\partial_{X^-}  \tilde{\phi})^2   \Big) {dX^-}^2 \Big].
\end{split}
\end{gather}

It could be seen that the difference between this induced metric and the corresponding one for the AdS case, eq. \ref{eq:inducedAdS} is much less than their actual metrics.

The extrinsic curvature on this boundary would be
\begin{gather}
K=L^{-1}_{WAdS}\  .\left(2+\frac{3L_{WAdS}}{r}-\epsilon^2 e^{-2 \tilde{\phi}}  \left(\frac{1}{r^2 \alpha^2 } \partial^2_{X^+} \tilde{\phi}+\frac{1}{r^4 (\alpha^2-1) } \partial^2_{X^-} \tilde{\phi}  \right) \right).
\end{gather}

One could compare this with the corresponding term for the AdS metric in \ref{AdSextrinsic}.

The metric of warped $\text{CFT}_2$ on the boundary is
\begin{gather}
ds^2=-e^{2 \rho (t^+, t^-) } \Big ( dt^+ dt^- - h(t^+, t^-) (dt^+)^2  \Big).
\end{gather}

To go from $(X^+, X^-)$ to $(t^+, t^-)$ coordinates, one could find the corresponding Jacobian $dX^+ dX^-= J dt^+ dt^-$ as 
\begin{gather}
\frac{i}{2} e^{2\rho (t^+, t^-) }=J\ . \ \frac{e^{2\phi}}{\epsilon^2}  \sqrt {r^6 \alpha^2 (\alpha^2-1)+\epsilon^2 e^{-2 \tilde{\phi} }  \left(r^4(\alpha^2-1) (\partial_{X^+} \tilde{\phi} )^2+r^2 \alpha^2 (\partial_{X^-} \tilde{\phi} )^2     \right)   },
\end{gather}

The above relation could then give a hint to derive a DBI action of WCFT case.

The final action that could be derived from this calculation would be
\begin{gather}
I_G=-\frac{L}{8\pi G_N} \int dt^+ dt^- \left [ (1+\frac{3L}{r} ) \Big((\partial _{t^+} \phi )^2+(\partial_{t^-} \phi )^2\Big) +r^6 \alpha^2 (\alpha^2-1) (2+\frac{6L}{r} ) e^{2\phi}   \right].
\end{gather}

The first term is the deformed kinetic term and the second term is the deformed potential term.

However, this relation is not exactly correct, since we used the Einstein gravity along the way, which does not have enough fields to produce the needed IR behavior and the exact warped AdS geometry as we want. Another action that we could consider could be the one shown in relation \ref{eq:lowerSpinGravity},  or other higher derivative gravity theories which have the warped AdS as their solutions such as Chern-Simons, New or Topologically Massive Gravity (TMG).

Here, we use the TMG for the gravity action. In the first order formalism, its action with a negative cosmological constant $\Lambda= - 1/\ell^2$ could be written as \cite{Miskovic:2009kr}
\begin{gather}\label{eq:TMGaction}
I= - \frac{1}{ 16 \pi G} \int_M \epsilon_{ABC} \left ( R^{AB} + \frac{1}{3 \ell^2} e^A e^B \right) e^C+ \frac{1}{32 \pi G \mu} \int_M \left( L_{CS} \left (\omega \right) + 2 \lambda_A T^A \right)+\int _{\partial M} B.
\end{gather} 

In the above action, $M$ is a three-dimensional manifold where $x^\mu$ are the local coordinates, $G$ is the gravitation constant, $\mu$ is a constant parameter with the dimension of mass and $L_{CS}$ is the gravitational Chern-Simons 3-form with the following relation
\begin{gather}
L_{CS} (\omega) = \omega ^{AB} d \omega_{BA}+\frac{2}{3} {\omega^ A}_B {\omega^B}_C {\omega^C}_A.
\end{gather}

By defining the dreibein $e^A= e^A_\mu dx^\mu $ and the spin connections $ \omega^{AB} =\omega_\mu ^{AB} dx^\mu $ one could write the curvature 2-form as
\begin{gather}
 R^{AB} =\frac{1}{2} R_{\mu\nu}^{AB} dx^\mu dx^\nu = d \omega^{AB}+ {\omega^A}_C \omega^{CB}. 
\end{gather}

For the TMG case, in \cite{Miskovic:2009kr}, the boundary term which makes the variational principle well-defined were introduced as 
\begin{gather} \label{eq:boundary}
B= \frac{1}{32\pi G} \epsilon_{ABC} \omega^{AB} e^C.
\end{gather}
This action then could be used after finding the Jacobians.

In topological and also Chern-Simons theories, the contribution of the boundary term is its most significant part as it is the case for the modes on the boundary of topological matters. Therefore, the gates dual to the boundary part play the most significant role.

One could also consider a parity-preserving theory such as New Massive Gravity (NMG) which contains warped AdS as its solutions, and then compare the result with the previous cases. The path-integral optimization complexity and the kind of quantum gates or circuits needed would take a different form in that case.

To review that theory, one could note that the action of NMG is
\begin{flalign}\label{eq:NMGaction}
I=\frac{1}{16 \pi G} \int_{\mathcal{M}} d^3 x \sqrt{-g} \left[ R-2\Lambda-\frac{1}{m^2} \left( R^{\mu \nu} R_{\mu \nu} -\frac{3}{8} R^2 \right) \right],
\end{flalign}
where $m$ is a dimensionful parameter. 

This theory could also be written in the following form \cite{Hohm:2010jc}
\begin{gather}
I=\frac{1}{16 \pi G} \int_{\mathcal{M}} d^3 x \sqrt{-g} \left[ R-2\Lambda+f^{\mu \nu} G_{\mu \nu}+\frac{m^2}{4} \left( f^{\mu \nu} f_{\mu \nu} -f^2 \right) \right ],
\end{gather}
where in the above term, $G_{\mu \nu}$ is the Einstein tensor and the auxiliary field $f_{\mu \nu}$ is
\begin{gather}
f_{\mu \nu}=-\frac{2}{m^2} \left( R_{\mu \nu}-\frac{1}{2 (d+1)} R g_{\mu \nu} \right ).
\end{gather}

The Gibbons-Hawking boundary term would be
\begin{gather}\label{eq:NMGboundary}
I_{GGH}= \frac{1}{16 \pi G} \int_{\partial \mathcal{M}} d^2 x \sqrt{- \gamma} \left( -2K-\hat{f}^{ij} K_{ij}+\hat{f} K \right),
\end{gather}
where $K_{ij}$ is the extrinsic curvature of the boundary and $K=\gamma^{ij} K_{ij}$ is the trace of the extrinsic curvature.

 The auxiliary filed $\hat{f}^{ij}$ is also defined as
\begin{gather}
\hat{f}^{ij}= f^{ij}+2h^{(i} N^{j)}+s N^i N^j.
\end{gather}

The functions in the above relation are defined from the following ADM metric
\begin{gather}
ds^2= N^2 dr^2+\gamma_{ij}( dx^i+N^i dr) (dx^j+N^j dr).
\end{gather}

In the NMG action, the mass of graviton plays the essential role which affects various quantum information quantities of the model, such as entanglement and its evolutions during quench processes, entanglement and complexity of purification, etc, which were studied in more details in \cite{Zhou:2019jlh,Ghodrati:2019hnn, Ghodrati:2018hss, Ghodrati:2017roz}. 

Since this action is parity-preserving, one would expect to see a final simpler tensor network structure and quantum circuit model. This point has been validated by constructing and comparing the Hawking-Page phase diagrams for both NMG and TMG warped black hole solutions in \cite{Ghodrati:2016ggy,Ghodrati:2016tdy,Ghodrati:2016vvf,Detournay:2015ysa}.

The effect of corner terms in the form of \cite{Hayward:1993my}
\begin{gather}
I_c=\frac{1}{16 \pi G_N} \int_\Sigma \sqrt{\gamma} (2\alpha-\pi).
\end{gather}
should also be considered at the end as it could affect the form of the circuit complexity and therefore the corresponding final tensor network model.

\subsection{Fibration of $\text{AdS}_3$ and deriving warped $\text{AdS}_3$}\label{sec:Fibration}

Before deriving a chiral, deformed Liouville action starting from a warped $\text{AdS}_3$ and using TMG, it is better to understand the components of warped AdS geometry, its fibration form and its Killing vectors.

The metric of AdS could be written as a spacelike (or timelike) fibration with fiber coordinate $u$ or $\tau$ as
\begin{gather}\label{eq:AdSMetric}
ds^2=\frac{\ell^2}{4} [-\cosh^2 \sigma d\tau^2+d\sigma^2+ (du+\sinh \sigma d \tau)^2 ]\\
=\frac{\ell^2}{4} [\cosh^2 \sigma du^2 +d\sigma^2-(d\tau+\sinh \sigma du)^2],
\end{gather}
which has $SL(2,R)_L \times SL(2,R)_R$ isometries. By multiplying the fiber by a warping factor, one could break the isometry group to $SL(2,R) \times U(1)$ and derive several solutions of warped $\text{AdS}_3$ as in \cite{Anninos:2008fx}.

In addition, in \cite{Anninos:2008fx}, the Killing vectors for various warped $\text{AdS}_3$ metrics have been derived which would be useful in studying the boundary CFTs and the induced metric on it. As it has been shown, the $SL(2, \mathbb{R} _L)$ isometries are given by
\begin{align}
J_1 & =\frac{2 \sinh u}{\cosh \sigma} \partial_\tau-2 \cosh u \partial _\sigma+2 \tanh \sigma \sinh u \partial_u, \nonumber\\
J_2 & =2\partial_u, \nonumber\\
J_0  & =-\frac{2 \cosh u}{ \cosh \sigma} \partial_\tau+2 \sinh u \partial_\sigma-2 \tanh \sigma \cosh u \partial_u .
\end{align}

Also, the isometries of $SL(2, \mathbb{R})_R$ are
\begin{align}
\tilde{J}_1 & = 2 \sin \tau \tanh \sigma \partial_\tau-2 \cos \tau \partial_\sigma+\frac{2 \sin \tau}{\cosh \sigma} \partial_u, \nonumber\\
\tilde{J}_2 & =- 2 \cos \tau \tanh \sigma \partial_\tau- 2 \sin \tau \partial_\sigma - \frac{2 \cos \tau }{\cosh \sigma} \partial_u, \nonumber\\
\tilde{J}_0 & = 2 \partial_\tau.
\end{align}

The point here is that, for the spacelike warped anti-de Sitter case, the Killing vectors are given by the $SL(2, \mathbb{R})_R$ and $J_2$, so for this case we should choose $u$ as the radial coordinate. For the timelike warped anti-de Sitter case, however, the Killing vectors are given by the $SL(2, \mathbb{R})_L$ and $\tilde{J}_0$, so for this case we choose $\tau$ as the radial coordinate. For the null warped AdS, the Killing vectors would be
\begin{gather}
N_1=\partial_-, \ \ \ \ N_0= x^- \partial_-+\frac{u}{2} \partial_u, \ \ \ \ N_{-1}=(x^-)^2 \partial_--u^2 \partial_+ +x^- u \partial_u, \ \ \ \ N= \partial_+.
\end{gather} 
For this case then $u$ should be chosen as the radial coordinate.

\subsubsection{Spacelike Warped $\text{AdS}_3$}
In this part, we start from a spacelike warped AdS metric and using TMG and the procedure introduced in \cite{Takayanagi:2018pml}, we derive a form of a deformed chiral Liouville action and then we study its properties.

The \textit{spacelike} or hyperbolic warped anti-de Sitter solution could be written as
\begin{gather}
ds^2=\frac{\ell^2}{(\nu^2+3)} \left[ -\cosh^2 \sigma d \tau^2+ d\sigma^2+ \frac{4 \nu^2}{\nu^2+3} (du+ \sinh \sigma d \tau)^2  \right].
\end{gather}

 The case of $\nu^2 >1$, is spacelike \textit{stretched} $\text{AdS}_3$, and the case of $\nu^2 <1$ is spacelike \textit{squashed} $\text{AdS}_3$.
 
 Note that near the boundary of spacelike warped $\text{AdS}_3$, where $ u \to \infty$ we could get the following geometry
 \begin{gather}
 \frac{ds^2}{ (\frac{2\nu}{\nu^2+3} \ell)^2 } \simeq   \left(\sinh ^2 \sigma- \cosh^2 \sigma \big( \frac{\nu^2+3}{4 \nu^2}\big) \right) d \tau^2 +\frac{\nu^2+3}{4\nu^2} d\sigma^2.
 \end{gather}
 
In this limit we could remove the non-diagonal term $2 \sinh \sigma du d\tau$ as well as the part $du^2$. This is the warped analogue of conformal transformation of the flat spacetime. This point to the fact that the warped CFT vacuum could be defined by the path-integral, and the two surfaces of $M_\Sigma^{(1)}$ and $M_\Sigma ^{(2)} $ in the bulk warped $\text{AdS}_3$ could be related by the Weyl and boost transformations in the UV region. 
 
 The lattice spacing then is determined by the coordinate $u$ and the lattice site corresponds to the unit area measured by $ ds^2/ (\frac{2\nu}{\nu^2+3} \ell)^2 $. In the warped AdS case, similar to the case of \cite{Takayanagi:2018pml} for AdS, one could deform the shape of each surface in the bulk using path-integral optimization and create a duality between the surfaces in warped AdS and the non-unitary quantum circuits of path-integration with an appropriate UV cut off which we will discuss further below.

So, in the bulk side, we consider the action
\begin{equation}\label{eq:GeneralAction}
\begin{split}
I_{\text{TMG}}&=\frac{1}{16\pi G} \int_{\mathcal{M}} d\sigma  \ du   \ d\tau \ \sqrt{-g} \left( R +\frac{2}{\ell^2} \right) \nonumber\\
&+ \frac{\ell}{96 \pi G \nu} \int_{\mathcal{M} } d\sigma \ du    \ d\tau \sqrt{-g} \epsilon^{\lambda \mu \nu} \Gamma^r_{\lambda \sigma} \left (\partial_\mu \Gamma^\sigma_{r \nu} +\frac{2}{3}\Gamma^\sigma_{\mu \tau} \Gamma^\tau_{\nu r} \right) \nonumber\\
&+ \frac{1}{32 \pi G}  \int_{\partial \mathcal{M}} du \ d\tau \ \epsilon_{ABC} \omega^{AB} e^C,
\end{split}
\end{equation}
and then calculating each term of this action for the spacelike warped AdS and then by performing the coordinate transformation, we could get a new gravity theory which corresponds to the complexity functional of path-integral optimization for the warped AdS.

The contribution from the first term is
\begin{gather}
\frac{1}{16\pi G} \int_{\mathcal{M}} d\sigma  \ du   \ d\tau \ \sqrt{-g} \left( R +\frac{2}{\ell^2} \right)=-\frac{\nu \ell (3\nu^2+8) }{4\pi G (\nu^2+3)^2} \int du d\sigma d \tau \cosh \sigma, 
\end{gather}
as the even part of the first term is
\begin{gather}
 \epsilon^{\lambda \mu \nu} \Gamma^r_{\lambda \sigma} \partial_\mu \Gamma^\sigma_{r \nu} (\text{even}) =\frac{4 \nu^2 (2\nu^2-3)}{(\nu^2+3)^2} \cosh \sigma , 
\end{gather}
and the odd part of the first term would give
\begin{gather}
 \epsilon^{\lambda \mu \nu} \Gamma^r_{\lambda \sigma} \partial_\mu \Gamma^\sigma_{r \nu} (\text{odd}) =\frac{4\nu^4 \cosh \sigma}{(\nu^2+3)^2}, 
\end{gather}
leading to
\begin{gather}
\epsilon^{\lambda \mu \nu} \Gamma^r_{\lambda \sigma} \partial_\mu \Gamma^\sigma_{r \nu}=\frac{12\nu^2 (\nu^2-1)}{(\nu^2+3)^2} \cosh \sigma.
\end{gather}

The even part of the second term in above relation gives
\begin{gather}
I_{CS1,\text{Even part}} = \frac{2\nu^2}{(\nu^2+3)^2} \sech \sigma \left (3+(2\nu^2-3) \cosh 2\sigma \right),
\end{gather}
and the odd part is just the negative of the above result. So for the second term we just get zero,
\begin{gather}
 \frac{2}{3} \epsilon^{\lambda \mu \nu} \Gamma^r_{\lambda \sigma}\Gamma^\sigma_{\mu \tau} \Gamma^\tau_{\nu r} =0.
\end{gather}

The total integral of the second line would then be
\begin{gather}
I_{CS}= \frac{\ell}{96 \pi G \nu} \int_{\mathcal{M} } d\sigma \ du    \ d\tau \sqrt{-g} \epsilon^{\lambda \mu \nu} \Gamma^r_{\lambda \sigma} \left (\partial_\mu \Gamma^\sigma_{r \nu} +\frac{2}{3}\Gamma^\sigma_{\mu \tau} \Gamma^\tau_{\nu r} \right) = \frac{\ell^4 \nu^2 (\nu^2-1)}{4\pi G (\nu^2+3)^4} \int d\sigma du d\tau \cosh \sigma.
\end{gather}

For calculating the last term, which is the boundary term, we choose the following vielbein and spin connections as
\begin{align*} 
e^0&=\frac{ \ell \sqrt{\left(7 \nu ^2-3\right) \sinh ^2\sigma -\nu ^2-3}}{\nu ^2+3}\text{$ d\tau $}, \nonumber\\ e^1&=\frac{ \ell}{\sqrt{\nu ^2+3}} \text{$ d \sigma $}, \nonumber\\ e^2&=\frac{2 \ell \nu}{\nu ^2+3}( \sinh \sigma \ \text{$d\tau $} +\text{$du$}),
\end{align*}
and
\begin{align*} 
\omega ^0{}_1&=\frac{\cosh \sigma  \Big( \left(5 \nu ^2-3\right) \sinh \sigma \text{ $ d\tau $} -2  \nu ^2 \text{$du$}\Big)}{\sqrt{\nu ^2+3} \sqrt{\left(7 \nu ^2-3\right) \sinh ^2\sigma -\nu ^2-3}}, \nonumber\\  {\omega^0}_2&=-\frac{\nu  \cosh \sigma }{\sqrt{\left(7 \nu ^2-3\right) \sinh ^2\sigma-\nu ^2-3}} \text{d$\sigma $} , \nonumber\\  {\omega^1}_2&=-\frac{ \nu  \cosh \sigma }{\sqrt{\nu ^2+3}} \text{d$\tau $}. 
\end{align*}

So we get

\begin{gather}
\epsilon_{ABC} \ \omega^{AB} e^C= \frac{-4\nu (7\nu^2-3)\cosh \sigma \sinh \sigma }{\ell \sqrt{\nu^2+3} \sqrt{(7\nu^2-3) \sinh^2 \sigma-\nu^2-3} }  du \wedge d\tau.
\end{gather}

The third term in the action then would give 
\begin{gather}\label{eq:totspaceaction}
\frac{1}{32 \pi G}  \int_{\partial \mathcal{M}} du \ d\tau \ \epsilon_{ABC} \ \omega^{AB} e^C=\int du d\tau \frac{-\nu (7\nu^2-3)\cosh \sigma \sinh \sigma }{8\pi G\ell \sqrt{ (\nu^2+3)}\sqrt{ (7\nu^2-3) \sinh^2 \sigma-\nu^2-3} },
\end{gather}

So the total action would be
\begin{gather}\label{eq:SpaceAction1}
I_G=  \frac{\nu \ell}{4\pi G} \int du d\tau \sinh \sigma \Big ( \frac{ \ell^3 \nu (\nu^2-1) }{ (\nu^2+3)^4 } -\frac{ 3\nu^2+8}{(\nu^2+3)^2} -\frac{(7\nu^2-3) \cosh \sigma }{2\ell^2 \sqrt{\nu^2+3} \sqrt{ (7\nu^2-3)\sinh^2 \sigma-\nu^2-3}} \Big).  
\end{gather}

Note that $\sigma$ here is a function of the two fields $\tilde{\rho}$ and $\tilde{h}$ which then could be written in terms of the actual $\rho$ and $h$ of the metric \ref{eq:chiralmetricgauge}, after the coordinate transformation. Therefore, the above action could lead to a deformed version of Liouville. To find the exact form in terms of $t^+, t^-, h, \rho$ after the coordinate transformation, we need to choose a suitable physical function for the \textit{``cut off surface"} for the warped geometries.

If we assume the following cutoff surface as
\begin{gather}
\sigma= -\epsilon \ . \ e^{- \tilde{\rho} (\tau,u)  } \Big(1- {\tilde{h}(\tau,u)} ^{-\frac{1}{2}} \Big),
\end{gather}
then after considering the chiral gauge relation, $\partial_-h=0$ \cite{Compere:2013aya}, which here would be $\partial_u \tilde{h}=0$, the induced metric on this boundary could written as
\begin{align}\label{eq:induced2}
ds^2 & =\frac{\ell^2 }{ (\nu^2+3) } \Big [ \Big ( \frac{3 \nu^2-3}{\nu^2+3} \sinh^2 \sigma-1+\epsilon^2 e^{-2 \tilde{\rho} } (\partial_\tau \tilde{\rho} -\partial_\tau \tilde{\rho} \  \tilde{h}^{-\frac{1}{2}} -\frac{1}{2} \tilde{h}^{-\frac{3}{2}} \partial_\tau \tilde{h})^2 \Big ) d\tau^2\nonumber\\& +
\Big( \frac{8 \nu^2  }{\nu^2+3} \sinh \sigma +2 \epsilon^2 e^{-2 \tilde{\rho} }  (\partial_\tau \tilde{\rho} -\partial_\tau \tilde{\rho} \ \tilde{h}^{-\frac{1}{2}} -\frac{1}{2} \tilde{h}^{-\frac{3}{2}}   \partial_\tau \tilde{h} )(\partial_u \tilde{\rho} -\partial_u \tilde{\rho} \ \tilde{h}^{-\frac{1}{2}} -\frac{1}{2} \tilde{h}^{-\frac{3}{2}} \partial_u \tilde{h}) \Big) du d\tau \nonumber\\ &+
\Big( \frac{4 \nu^2}{\nu^2+3} +\epsilon^2 e^{-2 \tilde{\rho} } ( \partial_u \tilde{\rho} - \partial_u \tilde{\rho} \ \tilde{h}^{-\frac{1}{2}} -\frac{1}{2} \tilde{h} ^{-\frac{3}{2} } \partial_u \tilde{h} )^2 \Big) du^2 \Big].
\end{align}

We then need to find the two Jacobians $J_\rho$ and $J_h$ which do the following transformations
\begin{gather}
du d\tau=J_\rho \ dt^+ dt^-,\ \ \ \ \ \
du d\tau=J_h \ dt^+ dt^+.
\end{gather}

For doing so, one should compare the relation \ref{eq:induced2}, with \ref{eq:WCFT1}.  Finding these Jacobians could then let us to write the action \ref{eq:SpaceAction1} in terms of $t^+, t^-, h, \rho$ which would lead to the deformed Liouville action in the ``actual coordinate".

\subsubsection{Timelike Warped $\text{AdS}_3$}
 The \textit{timelike} or elliptic warped anti-de Sitter solution could be found by warping the second part of \ref{eq:AdSMetric} as 
 \begin{gather}
 ds^2= \frac{\ell^2}{ (\nu^2+3)} \left [\cosh^2 \sigma du^2+d\sigma^2-\frac{4\nu^2}{\nu^2+3} (d\tau+\sinh \sigma du)^2  \right].
\end{gather}

The case of $\nu^2>1$ is timelike stretched and the case of $\nu^2<1$ is timelike squashed $\text{AdS}_3$. For the case of timelike stretched though we would have close timelike curves (CTCs).

Near the boundary of time-like warped $\text{AdS}_3$, where $ \tau \to \infty$ we have
 \begin{gather}
 \frac{ds^2}{ (\frac{2\nu}{\nu^2+3} \ell)^2 } \simeq   \left(-\sinh ^2 \sigma+ \cosh^2 \sigma \big( \frac{\nu^2+3}{4 \nu^2}\big) \right) d \tau^2 +\frac{\nu^2+3}{4\nu^2} d\sigma^2.
 \end{gather}
 
In this limit we also removed the term $-d\tau^2-2 \sinh \sigma  d\tau du$. 

The above metric is again the warped analogue of conformal transformation of the flat spacetime where timelike warped CFT vacuum is defined by the path-integral, so the two surfaces $M_\Sigma^{(1)}$ and $M_\Sigma ^{(2)} $ in the bulk time like warped $\text{AdS}_3$ could be related by Weyl and boost transformations in the UV region.

For the case where $M_\Sigma$ is timelike, the state $\ket{\Psi_\Sigma}$ could be obtained by a ``Lorentzian" path-integral on $M_\Sigma$. So the timelike surfaces of warped AdS could also be interpreted as quantum circuits.

For this case we should again calculate the action \ref{eq:GeneralAction}.

The contribution for each term is as follows. First,
\begin{gather}
\frac{1}{16\pi G} \int_{\mathcal{M}} d\sigma  \ du   \ d\tau \ \sqrt{-g} \left( R +\frac{2}{\ell^2} \right)=-\frac{\nu \ell (3\nu^2+8) }{4\pi G (\nu^2+3)^2} \int du d\sigma d \tau \cosh \sigma.
\end{gather}

The even part of the first term is
\begin{gather}
 \epsilon^{\lambda \mu \nu} \Gamma^r_{\lambda \sigma} \partial_\mu \Gamma^\sigma_{r \nu} (\text{symmetric}) =\frac{4 \nu^2 (2\nu^2-3)}{(\nu^2+3)^2} \cosh \sigma.
\end{gather}

The odd part of the first term would be
\begin{gather}
 \epsilon^{\lambda \mu \nu} \Gamma^r_{\lambda \sigma} \partial_\mu \Gamma^\sigma_{r \nu} (\text{non-symmetric}) =\frac{4\nu^4 \cosh \sigma}{(\nu^2+3)^2}.
\end{gather}

So we get
\begin{gather}
\epsilon^{\lambda \mu \nu} \Gamma^r_{\lambda \sigma} \partial_\mu \Gamma^\sigma_{r \nu}=\frac{12\nu^2 (\nu^2-1)}{(\nu^2+3)^2} \cosh \sigma.
\end{gather}

Then, for the second term we get
\begin{gather}
 \frac{2}{3} \epsilon^{\lambda \mu \nu} \Gamma^r_{\lambda \sigma}\Gamma^\sigma_{\mu \tau} \Gamma^\tau_{\nu r} =0.
\end{gather}

The even part of the above relation actually is
\begin{gather}
\text{even Part}= \frac{2\nu^2}{(\nu^2+3)^2} \sech \sigma \left (3+(2\nu^2-3) \cosh 2\sigma \right),
\end{gather}
and the odd part is just the negative of this result.

So the total integral of the second line would be
\begin{gather}
\frac{\ell}{96 \pi G \nu} \int_{\mathcal{M} } d\sigma \ du    \ d\tau \sqrt{-g} \epsilon^{\lambda \mu \nu} \Gamma^r_{\lambda \sigma} \left (\partial_\mu \Gamma^\sigma_{r \nu} +\frac{2}{3}\Gamma^\sigma_{\mu \tau} \Gamma^\tau_{\nu r} \right) = \frac{\ell^4 \nu^2 (\nu^2-1)}{4\pi G (\nu^2+3)^4} \int d\sigma du d\tau \cosh \sigma.
\end{gather}

For calculating the last boundary term, we choose the following vielbein and spin connections as
\begin{align*} 
e^0&=\frac{ 2 \ell \nu   \cosh \sigma}{\sqrt{\nu ^2+3} \sqrt{3 \left(1-\nu ^2\right) \sinh ^2 \sigma +\nu ^2+3}} \text{$ d\tau $}, \nonumber\\ e^1&=\frac{\ell}{\sqrt{\nu ^2+3}} \text{$ d\sigma $}, \nonumber\\ e^2&=\frac{ \ell\sqrt{3 \left(1-\nu ^2\right) \sinh ^2\sigma+\nu ^2+3}}{\nu ^2+3} \text{$du$} -\frac{ 4 \ell \nu ^2 \sinh \sigma }{\left(\nu ^2+3\right) \sqrt{3 \left(1-\nu ^2\right) \sinh ^2\sigma +\nu ^2+3}} \text{$ d \tau $},
\end{align*}
and
\begin{align*} 
{\omega^0}_1&= \frac{ \nu }{\left(\nu ^2+3\right) \sqrt{3 \left(1-\nu ^2\right) \sinh ^2\sigma+\nu ^2+3}} \big(4 \nu^2 \sinh \sigma d \tau+ (3 (\nu^2-1) \cosh 2 \sigma-\nu^2+9) du \big),\nonumber\\ \nonumber\\
{\omega^0}_2&=\frac{\nu  \left(3 \left(1-\nu ^2\right) \cosh (2 \sigma )+\nu ^2-9\right)}{\sqrt{\nu ^2+3} \left(3 \left(\nu ^2-1\right) \cosh (2 \sigma )-5 \nu ^2-3\right)} \text{$d\sigma $},\nonumber\\\nonumber\\
{\omega^1}_2&=\frac{ \cosh \sigma }{\sqrt{\nu ^2+3} \sqrt{3 \left(1-\nu ^2\right) \sinh ^2\sigma+\nu ^2+3}} \Big ( 2\nu^2 d\tau+3 (\nu^2-1) \sinh \sigma du \Big). 
\end{align*}

Therefore, the third term in the action would become
\begin{equation}
\begin{aligned}
&\frac{1}{32 \pi G}  \int_{\partial \mathcal{M}} du d\tau \epsilon_{ABC} \ \omega^{AB} e^C = \nonumber\\  &\frac{\nu}{16 \pi G \ell (\nu^2+3) } \int \Big( \sinh \sigma( -4\nu^2 + \frac{(\nu^2-3)(3+5\nu^2)-6\nu^2(\nu^2-1) \cosh(2\sigma) }{\nu^2+3+3(1-\nu^2) \sinh^2 \sigma} \Big) du d\tau. &
\end{aligned}
\end{equation}

So the total action could be written in a rather simplified form as
\begin{gather}\label{eq:SpaceAction}
I_G= \frac{\nu \ell}{4 \pi G} \int du d\tau \sinh \sigma \Big ( \frac{\ell^3 \nu (\nu^2-1) }{(\nu^2+3)^4}-\frac{3\nu^2+8}{(\nu^2+3)^2}-\frac{5 \nu^2+3}{4\ell^2 (\nu^2+3 +3 (1-\nu^2) \sinh^2 \sigma) } \Big).
\end{gather}

Note the difference between the last term in the above action and the corresponding one for the space-like case, in relation \ref{eq:totspaceaction}. For the undeformed case where $\nu=1$, the two actions would match.  Also, still here $\sigma (\tau, u) $ is a function of $u$ and $\tau$ which then needs to be integrated out. 

\begin{figure}[ht!]
\centering
\includegraphics[width=0.5\textwidth]{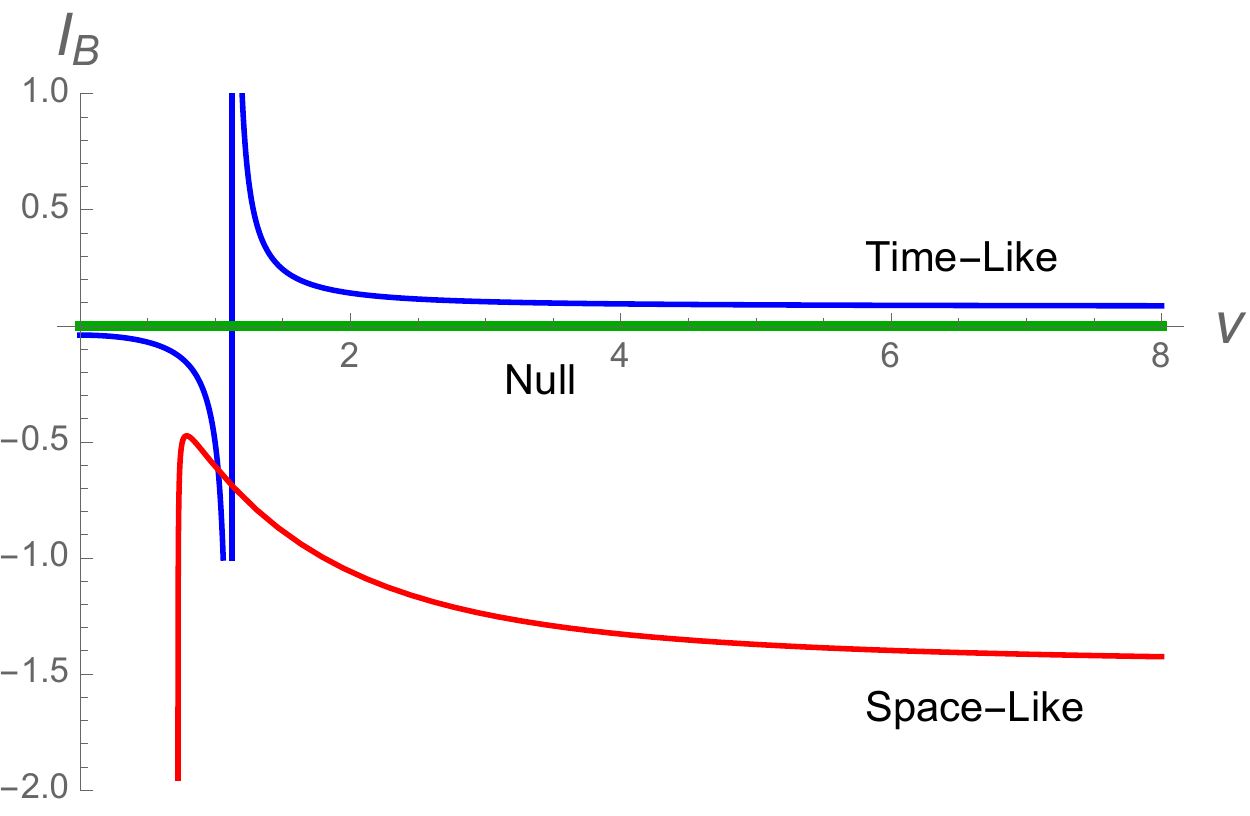}
\includegraphics[width=0.3\textwidth]{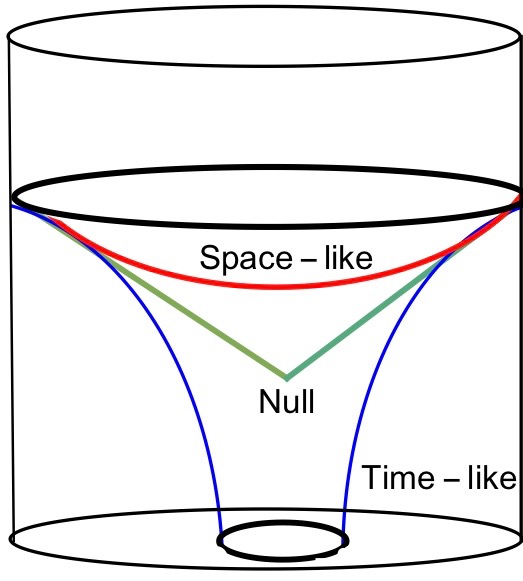}
\caption{ In the left part, the differentiating, boundary term in the chiral Liouville action versus deforming parameter $\nu$ for the case of time-like and space-like warped $\text{AdS}_3$ is shown. The right part is similar to figure 3 of \cite{Takayanagi:2018pml} which is for the case of AdS. The warped AdS would have deformed constructions depending on the value of $\nu$. These plots show the respective relations between these three manifolds }\label{fig:diffterm}
\end{figure}

Let's compare this differentiating term more closely. Assuming an arbitrary value for the $\sigma$ such as $\sigma=\frac{\pi}{2}$, and also $\ell=1$, we could compare this two terms in figure \ref{fig:diffterm}. There, one could see that, first, for a big enough value of $\nu$, the contribution from the boundary term for the time-like case is positive, and for space-like is negative and then when $\nu$ becomes very big, for both cases, they become constant. 

Also, one could see that for a specific value of the deformation, for the case of time-like or space-like geometries, the contributions to the action from the boundary term would blow up, indicating that at this value of $\nu$, the geometry would not be well-defined. This irregularity could also be modeled as the cosmological singularities similar to \cite{Das:2019cgl}.

This then even could be studied further from the point of view of \cite{Das:2019cgl}. For instance, a quench system in the warped AdS/CFT case could be considered and the fluctuations for this particular value of $\nu$ and the effects on the emergent space-time could be examined. For those special values of $\nu$, again one would expect that if the geometry is not well-defined, the couplings for fluctuations would diverge. Furthermore, one could look for the quench systems where a well-defined emergent warped AdS could be constructed.

Note that the case of $\nu<1$ leads to the squashed geometry and $\nu>1$ to the stretched case, while the case of $\nu=1$ corresponds to $\text{AdS}_3$ background. One could see that by increasing $\nu$, i.e. for the case of stretched AdS, the contribution to the action from the boundary term would be a decreasing function of $\nu$, while for the squashed case, the boundary action term for a specific value of $\nu<1$ would diverge.  For the case of null AdS, the boundary term would be zero. As a matter of fact, in Chern-Simons theory, most of the contributions to various physical parameters come from the boundary term. The consistency of these two pictures then once again shows that the ``complexity=action" could be used for the WAdS/WCFT case \cite{Ghodrati:2017roz, Takayanagi:2018pml}.

Then, to complete the calculations, we find the induced metric on the boundary
\begin{gather}
\sigma= -\epsilon \ . \ e^{- \tilde{\rho} (\tau,u)  } \Big(1- {\tilde{h}(\tau,u)} ^{-\frac{1}{2}} \Big),
\end{gather}
as

\begin{align}\label{eq:inducedT3}
ds^2 & =\frac{\ell^2 }{ (\nu^2+3) } \Big [\cosh^2 \sigma+\epsilon^2 e^{-2 \tilde{\rho} } (\partial_u \tilde{\rho} -\partial_u\tilde{\rho} \  \tilde{h}^{-\frac{1}{2}})^2 \Big ) du^2\nonumber\\& +
\Big( \frac{-8 \nu^2  }{\nu^2+3} \sinh \sigma +2 \epsilon^2 e^{-2 \tilde{\rho} }  (\partial_\tau \tilde{\rho} -\partial_\tau \tilde{\rho} \ \tilde{h}^{-\frac{1}{2}} -\frac{1}{2} \tilde{h}^{-\frac{3}{2}}   \partial_\tau \tilde{h} )(\partial_u \tilde{\rho} -\partial_u \tilde{\rho} \ \tilde{h}^{-\frac{1}{2}}) \Big) du d\tau \nonumber\\ &+
\Big(\frac{-4 \nu^2}{\nu^2+3} +\epsilon^2 e^{-2 \tilde{\rho} } ( \partial_\tau \tilde{\rho} - \partial_\tau \tilde{\rho} \ \tilde{h}^{-\frac{1}{2}} -\frac{1}{2} \tilde{h} ^{-\frac{3}{2} } \partial_\tau \tilde{h} )^2 \Big) d\tau^2 \Big],
\end{align}
where again the chiral gauge relation, $\partial_-h=0$ \cite{Compere:2013aya}, or $\partial_u \tilde{h}=0$ has been used.

 \subsubsection{Null Warped $\text{AdS}_3$}
 
The case of \textit{null} or parabolic warped $\text{AdS}_3$ would be
\begin{gather}
ds^2=\ell^2 \left [ \frac{du^2}{u^2}+\frac{dx^+ dx^-}{u^2} \pm \left(\frac{dx^-}{u^2} \right)^2 \right ].
\end{gather}

This metric is also a solution of TMG, but only for the case of $\nu^2= 1$. The pure $\text{AdS}_3$ could then be derived by removing the last term. 

It worths to mention that the null warped $\text{AdS}_3$ could be used to study cold atom systems \cite{Anninos:2008fx, Son:2008ye,Balasubramanian:2008dm}.

Again, for this case we calculate the action \ref{eq:GeneralAction},
\begin{equation}
\begin{split}
I_{\text{TMG}}&=\frac{1}{16\pi G} \int_{\mathcal{M}} du dx^+ dx^- \sqrt{-g} \left( R +\frac{2}{\ell^2} \right) \nonumber\\
&+ \frac{\ell}{96 \pi G \nu} \int_{\mathcal{M} } du dx^+ dx^- \sqrt{-g} \epsilon^{\lambda \mu \nu} \Gamma^r_{\lambda \sigma} \left (\partial_\mu \Gamma^\sigma_{r \nu} +\frac{2}{3}\Gamma^\sigma_{\mu \tau} \Gamma^\tau_{\nu r} \right) \nonumber\\
&+ \frac{1}{32 \pi G}  \int_{\partial \mathcal{M}} dx^+ dx^- \epsilon_{ABC} \omega^{AB} e^C.
\end{split}
\end{equation}

For this metric, considering the boundary being located at $u \ge \epsilon . e^{-\tilde {\rho} (x^+,  \ x^-) } $ would lead us to the effective boundary field theory.

The first part of the action would give
\begin{gather}
\frac{1}{16 \pi G} \int du dx^+ dx^- \sqrt{-g}  \left( R+\frac{2}{\ell^2} \right)=\frac{1}{16\pi G} \int du dx^+ dx^- \times \frac{\ell^4}{u^6}\nonumber\\
=\frac{\ell^4}{16 \pi G} \int dx^+ dx^- (-\frac{1}{5} u^{-5} )  \Big |_{u=\epsilon . e ^{-\tilde{\rho}(x^+, \  x^-) } }^\infty=\frac{\ell^4}{80 \pi G \epsilon^5} \int dx^+ dx^- e^{5 \tilde{\rho} (x^+, \ x^-) }.
\end{gather}

The Chern-Simons part would give

\begin{gather}
\frac{\ell}{96 \pi G \nu} \int_{\mathcal{M} } du dx^+ dx^- \sqrt{-g} \epsilon^{\lambda \mu \nu} \Gamma^r_{\lambda \sigma} (\partial_\mu \Gamma^\sigma_{r \nu} +\frac{2}{3} \Gamma^\sigma_{\mu \tau} \Gamma^\tau_{\nu r}  )=0.
\end{gather}

The even part has a factor of $\frac{4}{3u^3}$ and the odd part has a factor of $-\frac{4}{3u^3}$, which for the null warped AdS will sum up to zero.

For the null warped AdS, the tetrad and spin connections could be chosen as
\begin{gather}
e^0=\frac{\ell}{2} dx^+, \ \ \ \ e^1=\frac{\ell}{u} du, \ \ \ \ e^2=\frac{\ell}{2}dx^+ +\frac{\ell}{u^2}dx^-,\nonumber\\
\omega ^0{}_1=-\frac{\ell}{2} dx^+ -\frac{\ell}{u^2} dx^-,\ \ \ 
\omega ^0{}_2=-\frac{\ell}{u} du,\ \ \ 
\omega ^1{}_2=\frac{\ell}{2} dx^+ +\frac{2}{u^2} dx^-,
\end{gather}
or for another example, we could take
\begin{gather}
e^0=\frac{i \ell}{2} d x^+, \ \ \ e^1=\frac{\ell}{u} du, \ \ \ \ e^2=-\frac{i \ell}{2} dx^+ + \frac{i \ell}{u^2} dx^-,  \nonumber\\
\omega ^0{}_1=\frac{i}{2} \text{$d x^+ $}-\frac{i}{ u^2} \text{$dx^-$}, \ \ \  \ \ \omega ^0{}_2=-\frac{\text{$du $}}{u }, \ \  \  \ \ \omega ^1{}_2=-\frac{i}{2} \text{$ d x^+ $}+\frac{2 i }{ u^2} \text{$dx^-$},
\end{gather}
which using any of the two, would give a zero boundary term, i.e,
\begin{gather}
\epsilon_{ABC} \omega^{AB} e^C=0.
\end{gather}

The total action would be just
\begin{gather}
I_G=\frac{\ell^4}{80 \pi G \epsilon^5} \int dx^+ dx^- e^{5 \tilde {\rho}(x^+, x^-) }.
\end{gather}

Assuming $u=-\epsilon \ . \ e^{- \tilde{\rho} (x^+,x^-)} \Big(1- \tilde{h} (x^+,x^-)^{-\frac{1}{2}}  \Big) $, and then taking $\partial_{x^- } h=0$, the induced metric on the boundary would be

\begin{gather}
ds^2= \frac{\ell^2}{u^2} \Bigg \{\epsilon^2 e^{-2 \tilde{\rho} } \Big ( \partial_{x^+} \tilde{\rho} \big(1-\tilde{h}^{-\frac{1}{2}} \big)-\frac{1}{2} \tilde{h}^{-\frac{3}{2}} \partial_{x^+} \tilde{h} \Big)^2 {dx^+}^2 +
\Bigg ( \pm \frac{1}{u^2}+\epsilon^2 e^{-2 \tilde{\rho} } \Big (\partial_{x^-} \tilde{\rho} \big (1-\tilde{h}^{-\frac{1}{2}}\big )  \Big)^2 \Bigg) {dx^-}^2\nonumber\\+
\Bigg(1+\epsilon^2 e^{-2 \tilde{\rho}} \Big(\partial_{x^-} \tilde{\rho}  \big(1-\tilde{h}^{-\frac{1}{2} } \big)\Big) \Big(\partial_{x^+} \tilde{\rho}  \big(1-\tilde{h}^{-\frac{1}{2} } \big)-\frac{1}{2} \partial_{x^+} \tilde{h} \ \tilde{h}^{-\frac{3}{2}} \Big) dx^+ dx^- \Bigg \}.
\end{gather}

The null surfaces $M_\Sigma$ in the warped AdS could be understood as the degenerate limit of the time-like metrics. Also, the final result for complexity, similar to the AdS case, would be minimized for the null warped AdS geometries.

In this setup it would be possible to understand the way any component of the warped and deformed $\text{AdS}_3$ metric would emerge. It would also be possible to check how the density of unitary quantum gates, and the behavior of the scrambling of the quantum states of the dual deformed $\text{CFT}_2$ would be different from the usual CFT.

The same procedure could later be done for other geometries such as Lifshitz or hyperscaling violating \cite{Ghodrati:2014spa}, starting from the Newton-Cartan group theory, or the corresponding boundary algebra for flat or dS geometries or even by using other massive gravity theories; check \cite{Abedini:2019voz, Afshar:2015wjm,Afshar:2019npk} for more details of the dual algebra for such backgrounds.

\section{Holographic tools for warped conformal field theories}\label{sec:toolsWarp}

In this section we construct and extend various holographic tools for $\text{WAdS}_3/\text{WCFT}_2$ that have already been implemented in the setup of $\text{AdS}_3/\text{CFT}_2$  for studying the connections between geometry and encoded boundary information. 

\subsection{Tensor Network for WCFTs}\label{sec:TNWarped}

The idea of tensor network and renormalization of entanglement entropy to understand the mechanism of AdS/CFT and emergence of spacetime out of geometry has first started from the work of \cite{Swingle:2009bg}.

In \cite{Takayanagi:2018pml}, the holographic spacetimes have been modeled as quantum circuits of path-integrations. There, it has been argued that the holographic spacetimes are actually some collections of quantum circuits. Using path-integration method and an appropriate UV cutoff a codimension one surface in gravity has been connected to a quantum circuit model (extension of surface/state duality) and therefore the duality between tensor networks and AdS/CFT has been generalized.  There, the relations between the numbers of quantum gates to surface areas have been discussed and the holographic entanglement entropy formula has been generalized. In addition, the emergence of ``time" in AdS from the density of unitary quantum gates of CFT and also the gravitational force from quantum circuits has been shown. 

One could speculate about the indications of these results for the case of the warped or chiral geometries, for instance, how each deformation parameter or components of the metric, or even the presence CTCs of geometry are hidden in the boundary information. One could think that for these systems, other more ``exotic" quantum gates would be needed that one should search for.

In \cite{Milsted:2018vop},  the tensor networks as conformal transformation has been studied. There, in addition to \textit{disentanglers $u$} and \textit{isometries $w$} which construct the MERA, other forms of tensors have been studied. Examples are, \textit{euclideons $e$} which are the tensors in the euclidean time evolution and \textit{smoothers} which would be placed at the end of a truncated network.

Then, in \cite{Milsted:2018san}, \textit{eucliedons}, $e$ and \textit{lorentzions} $l$ have been defined. The authors assigned a path integral geometry to the Null, Euclidean and Lorentzian MERA. They found that the local rescaling in the CFT makes the path integral map $V$, while the tensor network map $\mathcal{W}$ remains unchanged. This rule which was explained further in \cite{Milsted:2018yur} could be applied for the warped CFT case as well.

\begin{figure}[ht!]
\centering
\includegraphics[width=0.8\textwidth]{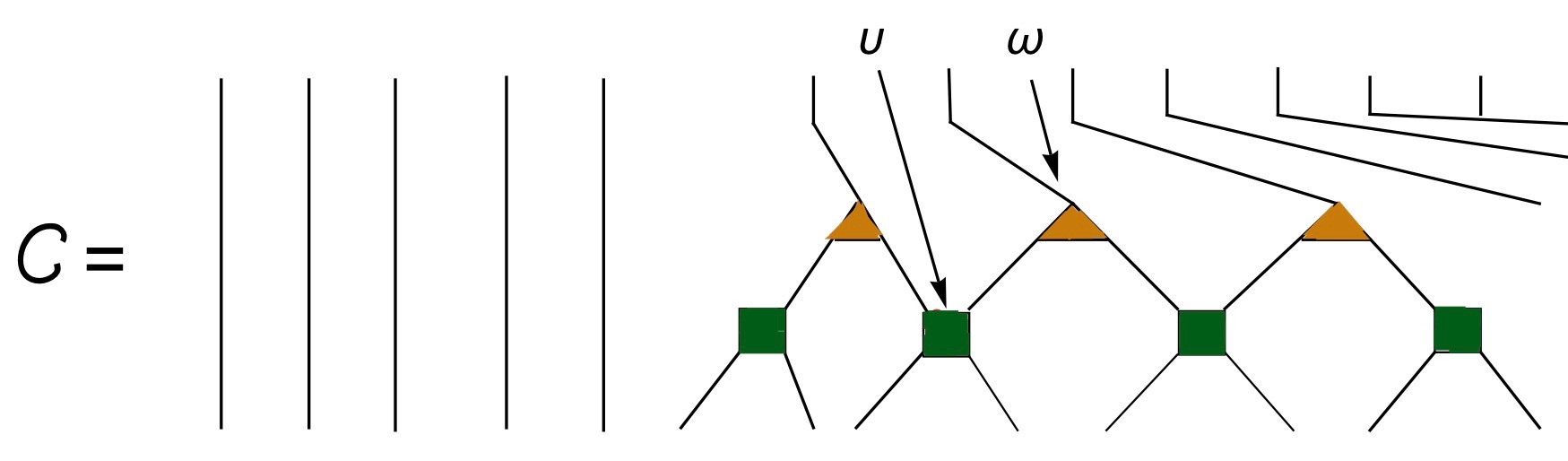}
\caption{The tensor \textit{chiraleon} for constructing WCFTs. Note that the yellow triangles are gates that are combination of unitary disentaglers and chiraleons. So one could consider the two layers are combined together.  }\label{fig:chiraloenexample}
\end{figure}

Here we would like to point out to several other features that a proposed tensor network for warped CFTs should poses. First, we noted that, for constructing tensor network for warped CFTs we actually need \textit{chiraleons} which produce only right moving dilation symmetry. This point has been considered to show a proposed idea for the structure of warped MERA as in figure \ref{fig:chiraloenexample}.  In fact, the WCFTs have a right moving stress tensor $T$ where $\partial_+ T=0$, and a right moving current $P$ where $\partial_+ P=0$. However, the right moving current generates left translations. So the transformations of the left-moving direction $x^+$ are generated by the right-moving current $P(x^-)$. Therefore, the insertion of chiraleons would be necessary.
 
Also, unlike the usual AdS/CFT systems, here instead of coupling to the Riemannian geometry which is the case of $2d$ CFT, the WCFTs should be coupled to the ``warped geometry" \cite{Hofman:2014loa} discussed before, which is a variant of the Newton-Cartan geometry.
 
In \cite{Jensen:2017tnb} it was argued that WCFTs have a dynamical critical exponent $z=\infty$, as under dilatations, time rescales but space does not. However, since for constructing the tensor network, we fix time, this factor would not affect our analysis.

\begin{figure}[ht!]
\centering
\includegraphics[width=0.8\textwidth]{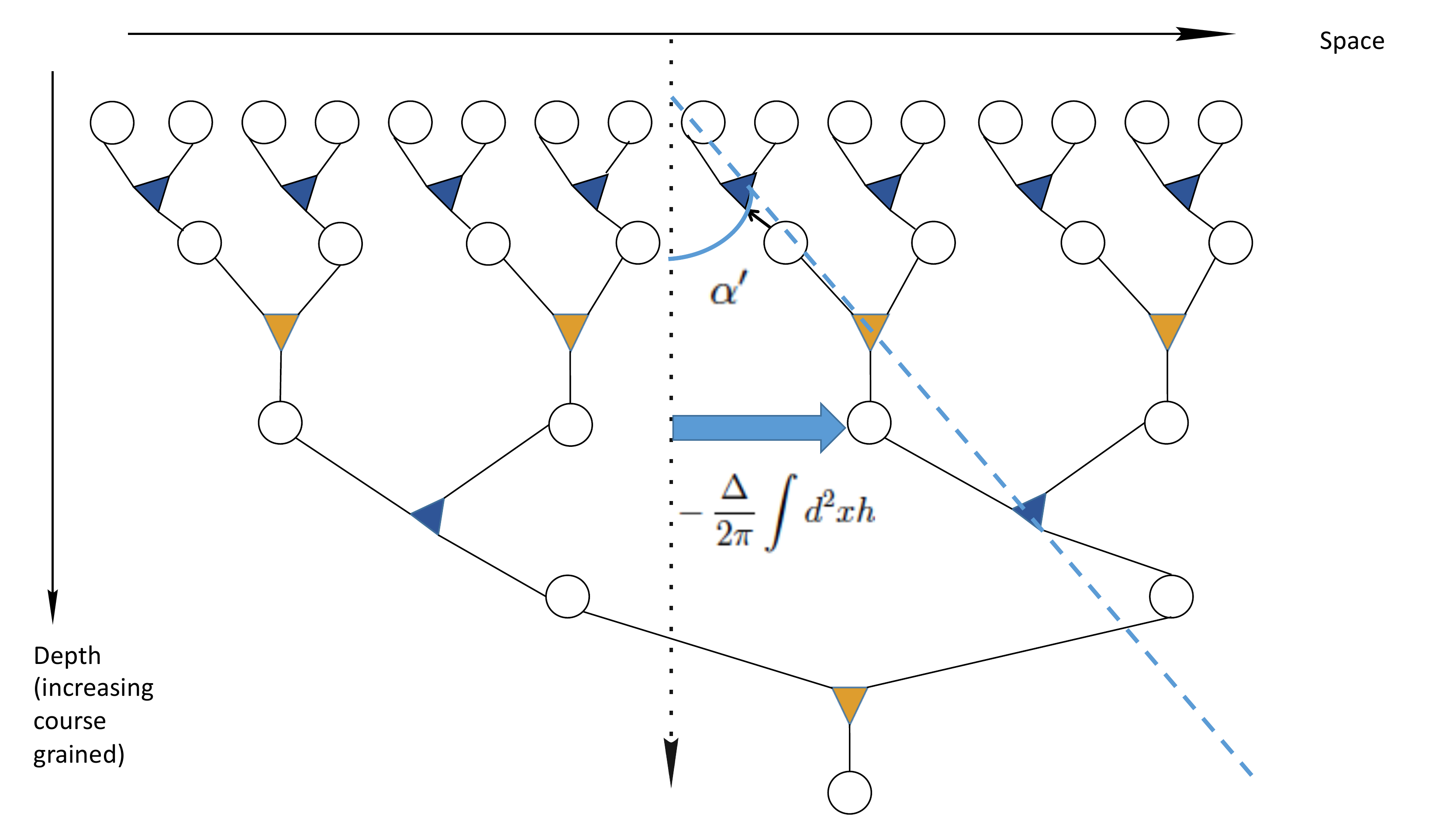}
\caption{The chiral structure of TN for warped CFT creates a deformed version of warped AdS space. The blue triangles could be a combination of chiraleons and disentanglers. Also, yellow triangles here are isometric coarse graining transformations and circles are the lattice sites. So this structure due to the presence of chiraleons does not have a dilation invariance in the left direction.}\label{fig:tensorWarpedParameters}
\end{figure}

In addition, insights from the constructions of Hawking-Page phase diagrams for the vacuum warped $\text{AdS}_3$ and warped AdS black holes which have already been established in \cite{Ghodrati:2016ggy, Detournay:2015ysa, Ghodrati:2017roz} could point out to the fact that the construction of geometry from the tensor network for WCFT should be \textit{``deformed"} in one direction as it has been shown in the schematic figure \ref{fig:tensorWarpedParameters}.

To understand better the difference between the tensor network system for the CFTs, versus those of WCFTs, we could examine further the difference between the Liouville action versus chiral Liouville gravity. 

 As we saw, the usual Liouville gravity theory deriving from the nonlocal Polyakov action written in the conformal gauge would be \cite{Compere:2013aya}
\begin{gather}
S^0_L= \frac{c}{96\pi} \int d^2x (Z \mathcal{R} -2 \lambda \sqrt{-g}),\ \ \ \ \ 
\mathcal{R} \equiv \sqrt{-g} R,\nonumber\\
Z(x) \equiv \int d^2 x' G(x,x') \mathcal{R} (x'),\ \ \ \ 
\sqrt{-g} g^{ab} \nabla_a \nabla_b G(x,x') = \delta^2 (x,x').
\end{gather}

On the other hand, in the case of the chiral Liouville gravity (Polyakov action in a chiral gauge), for studying the theory in a fixed sector for the left moving zero modes, an additional term in the following form should be added to the $S^0_L$ term, as
\begin{gather}
S_L=S^0_L+\frac{\Delta}{4\pi} \int d^x \sqrt{-g} g^{--}=S^0_L-\frac{\Delta}{2\pi} \int d^2 x h,
\end{gather}
which makes all the difference. Later, the effects of this additional term on the complexity from the view of complexity $=$ action should also be considered further.

One could then see that the action becomes asymmetric and the deformation at each point of the tensor network becomes a function of the parameters $h(t^+, t^-)$ where the left moving energy density is $\Delta/2$.

In the canonical formalism of chiral Liouville gravity, $t^-$ could then be considered as time. Then, from the relation $\partial_- h=0$, one could see that $h$ becomes independent of time which then is a necessary point in constructing the tensor network of WCFTs.

To get further information, similar to \cite{Witteveen:2019lsk}, using the wavelet decompositions, the quantum circuits for Majorana fermions could also be constructed. Similar to the Schrödinger holography, for the case of WAdS/WCFT, the Fourier modes around the spatial circle of $\text{WAdS}_3$ with momentum $k$ are dual to the boundary operators whose conformal dimension depends on $k$.

Other ideas such as quantum error correction could also help to get a better picture. In \cite{Camargo:2019isp}, where the connections between path integral complexity and circuit complexity has been outlined, the authors showed that the Liouville action which in general is expressed in terms of Weyl parameter, could be considered as a general cost function in the gate counting setup. In terms of quantum error correcting codes (QECCs), the dual of this physical CFT anomaly, have been conjectured to be a logical bulk gate. The same result still could be applied for WCFTs where we posses a ``warped Weyl" symmetry. Specifically, the level of the energy of the dual of these operators in terms of the eigenstate thermalization hypothesis  (ETH) \cite{Bao:2019bjp} have been studied, which could be repeated for WCFTs. One then could check whether for WCFTs the gates would satisfy the ETH ansatz. In addition, using this setup, one could estimate the energy difference for a layer of chiraleon as $\big | E_{i+1} - E_i \big | \propto  e^{-c N}$, where $c$ is an independent parameter related to the deformation parameter $\nu$ in the metric or $\alpha$ in the action.

Another point is that the complexity of Hamiltonian circuit $C(e^{-i t H})$ would be proportional to $\sqrt{-g_{tt}}$. For the warped $\text{AdS}_3$ case it is proportional to the deforming parameter $\alpha$. So for a fixed time period, the number of quantum gates in one direction relative to the number of quantum gates in the other direction would exactly determine the deforming parameter of the bulk geometry. It could in fact determine wether the geometry is squashed or stretched and also determine the extent of this deformation with respect to other parameters of the theory.  From the Hamiltonian formalism of chiral Liouville theory, we deduced that this deformation is actually related to the ``energy density parameter $\Delta$". This argument could then point out to a relation between the deformation parameter $\nu$ and the emergence of ``time" from the density structure of the deformed circuit in the WCFTs tensor network structure.

The coupling of the WCFTs to the background fields should also be considered, as this point could manifest itself in the causality wedge during the emergence of space from the tensor network. Here, instead of the light-cone in the CFT case, for the WCFT case we have another geometric structure called ``scaling structure" which essentially plays the same role. There, instead of the Weyl invariance we get a corresponding ``warped Weyl invariance" which should be used similarly.

\subsection{Entanglement structures in warped $\text{AdS}_3$} \label{sec:EEevolution}
In this section, we mention several observations about the structure of entanglement entropy (EE) of warped CFTs in several separate points, as for understanding better the renormalization group flow, structures of tensor network and the connections between chirality of the boundary theory and emergent geometry, EE and extremal surfaces still are very important tools.

For the case of WCFTs, the entanglement entropy has been studied in various works, see \cite{Castro:2015csg,Song:2016gtd,Song:2016pwx} for instance. 

The spacelike warped $\text{AdS}_3$ metric in global coordinates with warp factor $a \in [0,2) $ would be written as \cite{Anninos:2013nja}
\begin{gather}
ds^2= \frac{\ell^2}{4} \Big( -(1+r^2) d\tau^2 +\frac{d\tau^2}{1+r^2}+a^2 (du+r d\tau)^2 \Big).
\end{gather}

For this metric, similar to \cite{Czech:2017ryf}, starting from the derived entanglement entropy and the proposed model of tensor network for warped CFTs, by varying the circuit path integral-optimization complexity using the chiral Liouvile action, one could derive the equations of motion \ref{eq:eqofmotion} or \ref{eq:motionEQ}. 

In fact, the entanglement entropy of warped CFTs has been found in \cite{Castro:2015csg,Anninos:2013nja} and then modified in \cite{Song:2016gtd}. Later, we will see that we really need to use the modified version of it in order to get a physically well-behaved kinematic space.

The simpler relation which first we use to explain the structure of entanglement here is
\begin{gather} 
S_{EE}=-4L_0^{\text{vac}} \log \left( \frac{L}{\pi \epsilon} \sin \frac{\pi \ell}{L}\right)+i P_0^{\text{vac}} \ell \left( \frac{\bar{L}}{L}-\frac{\bar{\ell} }{\ell} \right),
\end{gather}
where $\ell$ and $\bar {\ell}$ are the separations of the endpoints of an assumed interval in space and time respectively,  and $L$ and $\bar {L}$ are related to the identification pattern of the circle which defines the vacuum of the theory. The second term would actually act as the twist term. The interval is considered to be on the plane, and then one assumes $L, \bar{L} \to \infty$. The angle $\alpha \equiv \frac{\bar{L}}{L} $ could also be defined, see figure \ref{wapredDomain}.

In \cite{Jiang:2019qvd}, the emergence of gravitation from entanglement in holographic CFTs which have a gravitational anomaly (with unbalanced left and right moving central charges) has been studied. The Wald-Tachikawa covariant phase space formalism has been implemented there in order to obtain the linearized equations of motion of TMG from the entanglement.

For the gravity action with Chern-Simons term, the extremization prescription, corresponding to Ryu-Takayanagi formula should then be modified in the following way
\begin{gather}
S_{HEE}= \text{ext}_\gamma \frac{1}{4G_N} \int_\gamma ds \left ( \sqrt{g_{\mu \nu} \dot{X}_\mu \dot{X}_\nu  }+\frac{1}{\mu} g_{\mu \nu} \tilde{n}^\mu \nu^\rho  \nabla_\rho n^\nu \right).
\end{gather}

This relation, for the locally warped $\text{AdS}_3$, could be considered as the length of the geodesic and the twist of the normal frame along the geodesic, as
\begin{gather}
S_{HEE}= \frac{\text{Length}}{4G_N}+\frac{\text{Twist} }{4 G_N \mu}.
\end{gather}

In the case of complexity of such theories also, a corresponding twist operator could be imagined which produces the specific warped anomaly, and then the total number of these operators create the additional term in the chiral Liouville action.

So for the warped AdS case, one could also have a corresponding relation of \cite{Takayanagi:2018pml} in the form
\begin{gather}
dS_{A \bar{A}}=\frac{dA (\Gamma_{P \bar{P} } ) }{4G_N}+\text{Twist}.
\end{gather}

\begin{figure}[ht!]
\centering
\includegraphics[width=0.5\textwidth]{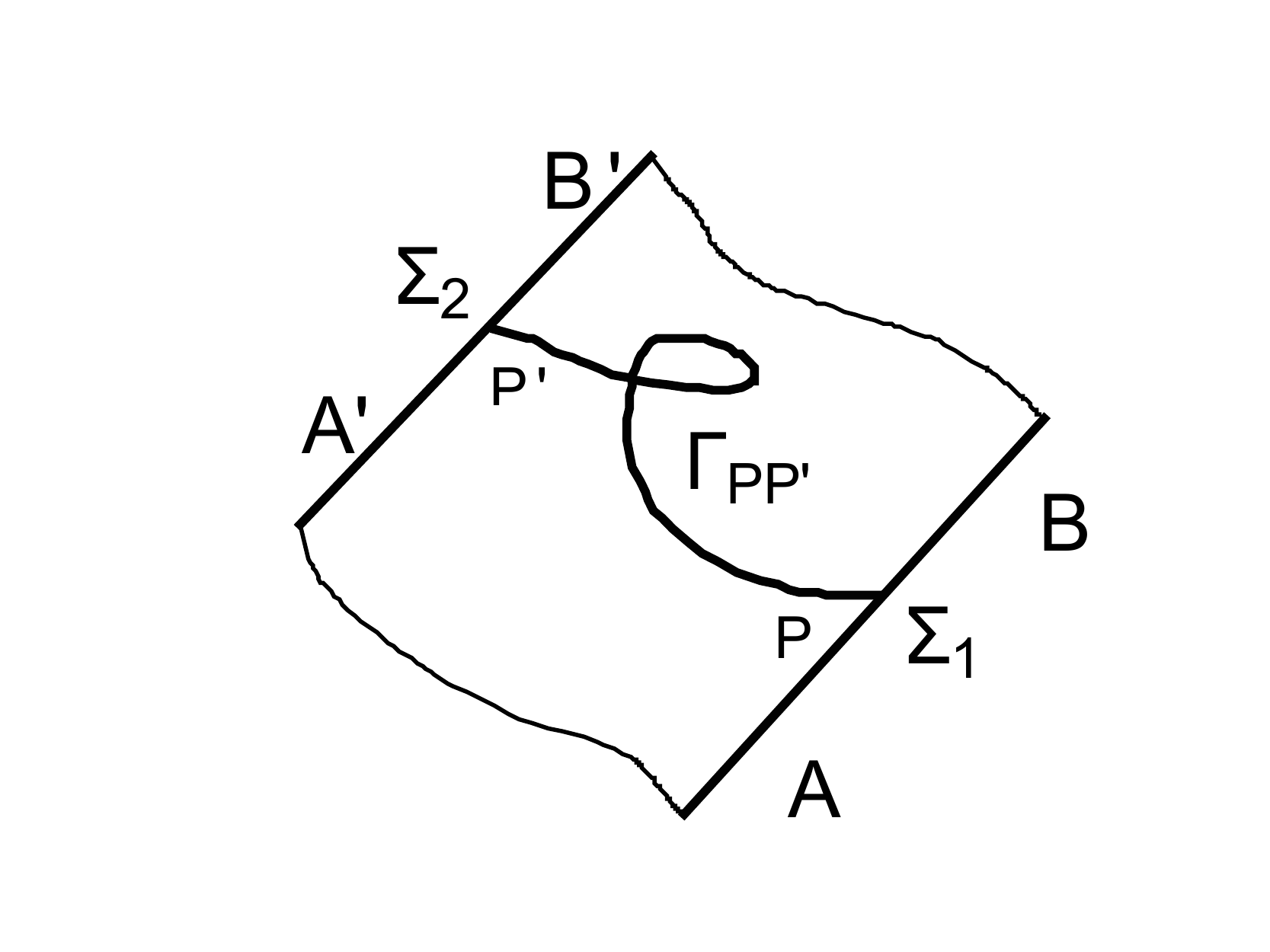}
\caption{ The codimension surface $\Gamma_{P \bar{P}}$ which connects $P$ and $\tilde{P}$ in the warped $\text{AdS}_3$.  }\label{warpedEEfig}
\end{figure}

For the warped CFTs, the evolution of entanglement entropy corresponds to the sum of the area elements and the twist. The deformation of the manifold by the Weyl anomaly of warped CFTs will then come into play which the schematic form has been shown in Fig. \ref{warpedEEfig}.

The twist fields of WCFT which would be the local operators in $\mathcal{C}^q$ are in the following form
\begin{gather}
\langle  \mathcal{O}(\Sigma^{(i) } ) \rangle_{\mathcal{C}, \mathcal{R}_q} =\frac{\langle \mathcal{O}_i (\Sigma) \Phi_q (\Sigma_1) \Phi_q^\dagger (\Sigma_2) \rangle_{\mathcal{C}^q, \mathbb{C}} }{ \langle \Phi_q (\Sigma_1) \Phi_q^\dagger (\Sigma_2) \rangle_{\mathcal{C}^q, \mathbb{C}}},
\end{gather}
where the conformal dimension and charge of these twist fields $\Phi_q$ have been found as
\begin{gather}\label{renyiQ}
\Delta_q= q \left ( \frac{c}{24}+\frac{L_0^{\text{vac} } }{q^2} \right), \ \ \ \ \ \ Q_q={P_0}^{\text{vac} }.
\end{gather}

Then, the effect of any non-trivial topology could be modeled by considering $q$ decoupled copies of field theories and then adding local twist fields $\Phi_q(\Sigma)$ at the endpoints of the interval. This would then couple together the replica copies.

Moreover, the complete partition function of warped CFTs has been found in \cite{Castro:2015csg} as
\begin{gather}
Z_{\bar{a} | a} (\bar{\tau} | \tau) = \exp \left (\frac{i \pi k}{2} (\frac{\bar{\tau}^2 a }{\tau} )+ 2 \pi i P_0^{\text{vac}} (\frac{a \bar{\tau}}{\tau} - \bar{a} )+\frac{2\pi i a}{\tau}L_0  \right )
\end{gather}

The complexity path integration and the replica manifold could then become connected to each other using the above relations. As mentioned in \cite{Castro:2015csg}, tthe OPE of the twist field in the correlation function of WCFTs with the $U(1)$ current $J(x) \Phi_q(y) \sim \frac{i  Q_q \mathcal{O} }{x-y} $, could also detect a singularity which shows its signature in the dual geometrical space.

The main point here then could be seen from taking the limit of $q \to 1$ in relation \ref{renyiQ}. The result would be non-zero due to the fact that the vacuum of the theory is not invariant under $SL(2, \mathbb{R}) \times U(1)$. Changing from cylinder to plane, the vacuum would be mapped to a non-trivial operator.  So the vacuum operator would be related to $\Phi_1$, which is being inserted locally at the end points of the interval, shown in figure \ref{warpedEEfig}. 

Also, similar to the case of entanglement entropy, the path-integral complexity is a function of vacuum charges $L_0^{\text{vac} }$ and $P_0^{vac}$, and so only the central charge of the theory would not be enough. As most realistic field theories, such as warped CFTs or non-unitary CFTs, are not invariant under conformal group, this statement is actually important and it should be applied while studying their circuit complexities.

\subsection{Surface/State Correspondence for $\text{WAdS}_3$/$\text{WCFT}_2$}\label{sec:sscorres}

The surface/state correspondence was introduced in \cite{Miyaji:2015yva}. If this indeed is a framework that could work for any quantum gravity theory; i.e, saying that tensor networks are really equivalent to gravitational theories, then one should consider how this framework could be applied for the chiral theories such as warped CFTs, and then check how a deformed AdS spacetime could be emerged by considering some particular distributions of quantum states in the chiral gauge. Then, one would like to understand how the quantum circuits of path integration in the setup of warped CFTs make the evolution of quantum entanglement, and how the area of a particular surface and the number of quantum gates creating entanglement could be connected. In this section we consider this problem. We also study the specific form of the information metric for WAdS/WCFT case.

As for the tensor network structure of warped CFTs, one could imagine that the number of intersections between any convex surface $\Sigma$ and the links between \textit{chiraleons} and other elements of the tensor network which intersect $\Sigma$ should determine the chirality of the theory. In addition all kinds of links between various gates on the network which intersects $\Sigma$ could give an estimation of the effective entropy $S(\Sigma) $ of the particular surface. However, as mentioned in \cite{Miyaji:2015yva}, this would not be a precise match as it just could act as a first estimation, since the links would not necessarily be maximally entangled and efficient. However, we could still take this first order estimation in what follows.

For the warped $\text{AdS}_3$ metric in the Poincare coordinate,
\begin{gather}\label{WAdS2metric}
ds^2=L^2_{WAdS} \left ( \frac{dr^2}{r^2}-r^2 {dX^-}^2+\alpha^2 (dX^+ + rdX^-)^2 \right),\nonumber\\  \ \ \ L_{WAdS}=\frac{\ell^2}{\nu^2+3},\  \ \ \ \ \ \alpha^2=\frac{4\nu^2}{\nu^2+3}, 
\end{gather}
one could compare the results we already got with those of \cite{Miyaji:2015yva}. 

\begin{figure}[ht!]\label{TNSEFF}
\centering
\includegraphics[width=0.35\textwidth]{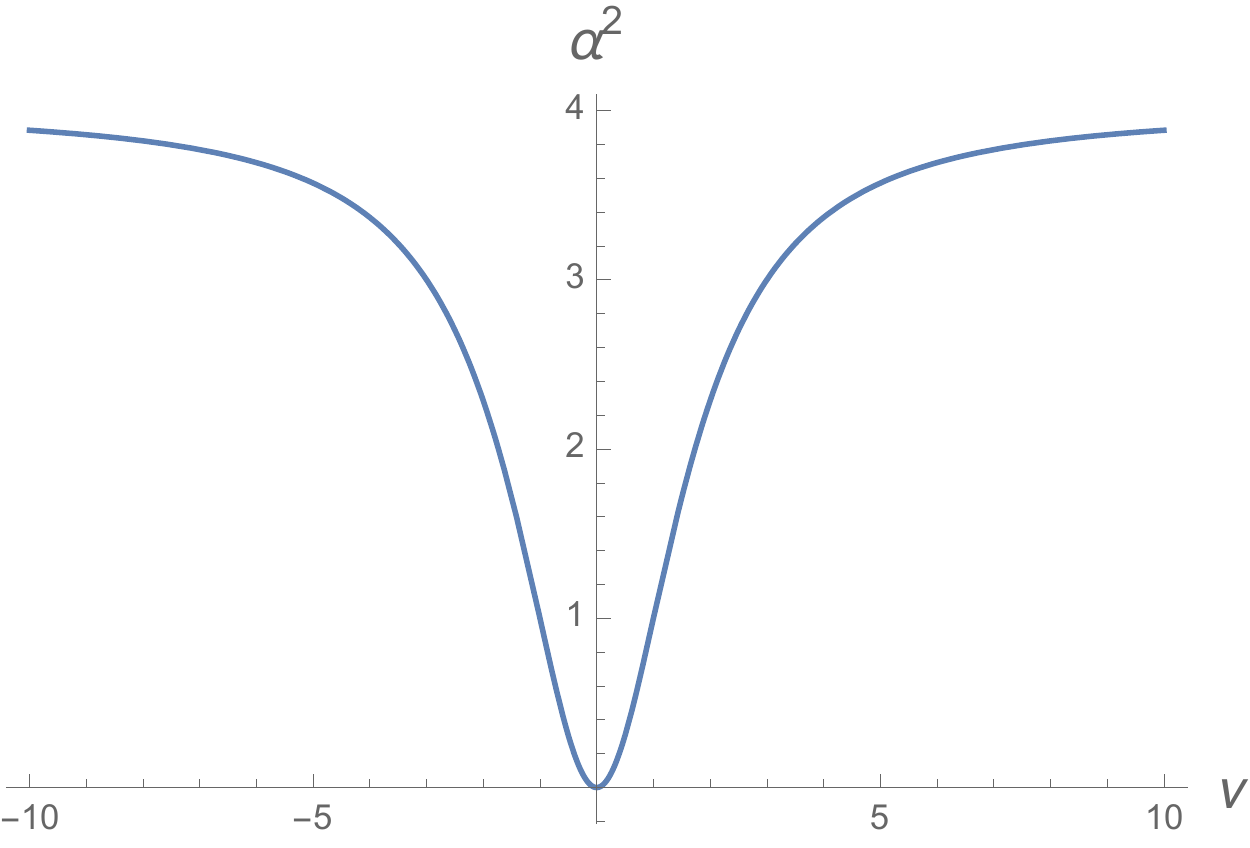}\hspace{15mm}
\includegraphics[width=0.35\textwidth]{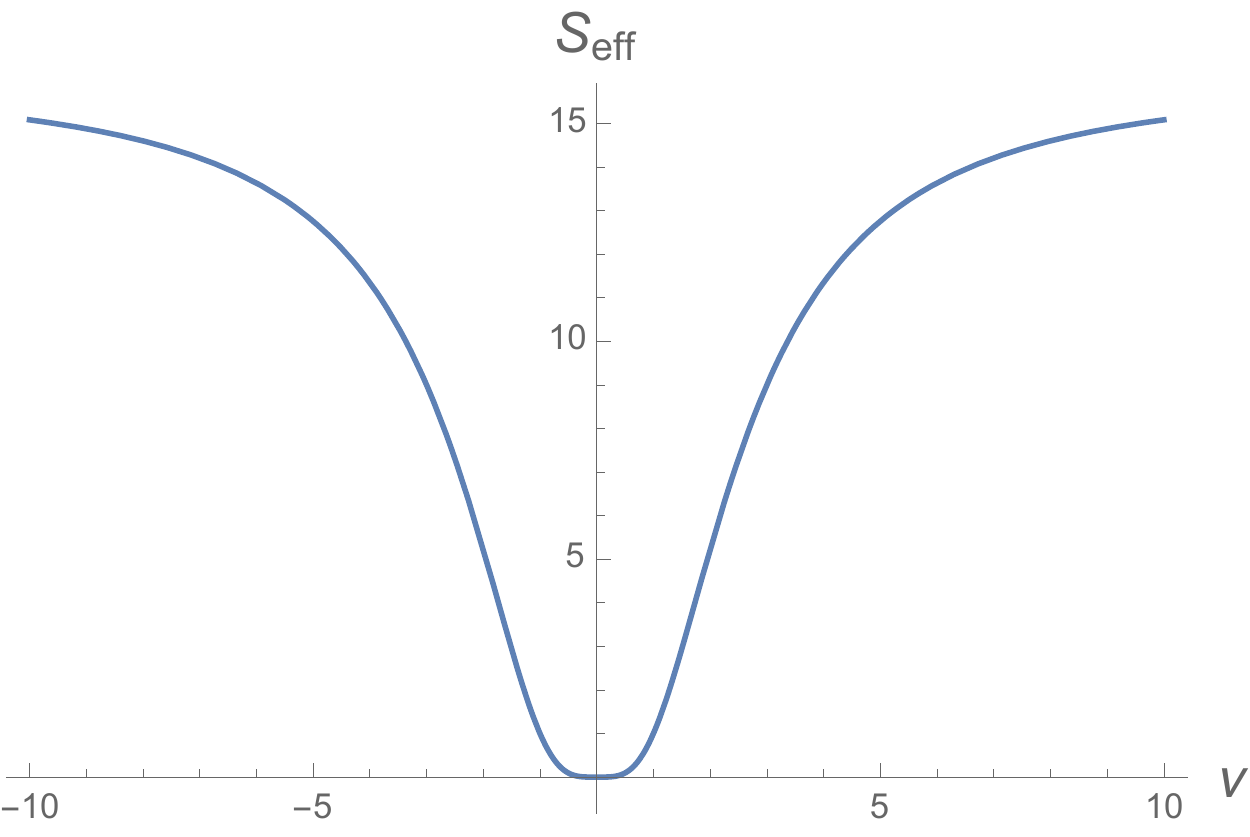}
\caption{The parameter $\alpha^2$ and the effective entropy $S_{\text{eff}}$ versus deforming factor $\nu$. The left plot could give a first estimation of the number of chiraleon gates versus the deformation of the geometry and the right one could show a c-like theorem for WCFTs as the effecting action is a monotonically increasing function of the physical positive values of $\nu$.}
\label{deformTN}
\end{figure}

First, as the deformation parameter $\alpha$ depends on the chirality and therefore the number of chiraleons intersecting the surface, one could check its effect on the geometry as well. The figure of $\alpha^2$ versus deforming parameter $\nu$ is shown in figure \ref{deformTN}. By warping the geometry further, some more chiraleon gates would be needed. Considering a positive $\nu$, it could be seen that the behavior is a monotonically increasing function of the deforming parameter, and then it becomes a constant value for larger $\nu$s, i.e, here it happens around $\nu \sim 4$.

In this case, the surface $M_\Sigma$, similar to the Lorentzian AdS \cite{Takayanagi:2018pml}, could be time-like, null or space-like. The null warped AdS (G$\ddot{\text{o}}$del geometries) could be considered as a degenerate limit of time-like surfaces. The space-like surfaces would then similarly correspond to the path-integrals on the space-like surfaces $M_\Sigma$. 

Path-integration then would change the normalization of the wave functional. The corner contribution in the gravity dual would actually point to the (non-Lorentzian) evolutions in these models. Also, one should note that similar to the AdS case, in WAdS geometries, the propagation of local excitations could break the causality in the bulk and therefore the circuits would be non-unitary. In fact, the quantum circuit on $M_\Sigma$ includes both unitary and non-unitary quantum gates.

If one considers the state dual to the vacuum state of warped CFT, i.e, $\ket{0}$ as $\ket{\Psi_\Sigma}$, then due to the warped Weyl invariance here and similar to the conformal scenarios, a time-like path-integration which starts from another quantum state $\ket{\Psi_{\tilde{\Sigma} } }$ would not affect modes that their wave lengths are larger than the ones in $\Sigma$. The time-like path integration would only create vacuum states for the modes whose lengths are between the ones in $\tilde{\Sigma}$ and $\Sigma$.

The Weyl invariance of path-integrations then points out to the fact that, similar to the AdS case, for any codimention one ``time-like" surface $M_\Sigma$ in warped $\text{AdS}_3$ which would connect the surface $\Sigma$ to $\tilde{\Sigma}$, the dual quantum circuit would also maps $\ket{\Psi_{\tilde{\Sigma} } }$ to $\ket{\Psi_{\Sigma}}$. So the codimension one surface $M_\Sigma$ could be regarded as a path-integration with a suitable cut off and then by a discretization with the cut off scale, this path integration could be regarded as a quantum circuit.

To complete our discussion here, similar to \cite{Miyaji:2015yva}, for the $\text{WAdS}_3$ geometries, the Fisher information metric $G_{uu}^{(B)}$ could be calculated. The definition of this metric is as follows
\begin{gather}
1- \big | \langle \Phi(\Sigma_u) \big | \Phi( \Sigma_{u+du} ) \rangle \big | = G_{uu}^{(B)} du^2.
\end{gather}

In \cite{Miyaji:2015yva}, in their eq. (3.19),  a conjectured from for $G_{uu}^{(B)}$ for the AdS case and in a gravity theory in $M_{d+2}$ has been proposed. For the warped AdS metric of \ref{WAdS2metric}, as the determinant of the non-radial parts $X^+$ and $X^-$ is $\sqrt{g}=\alpha r $, the above relation would result in the following form of Fisher metric
\begin{flalign}\label{infmetricWADS}
G_{uu}^{(B)}&=\frac{1}{G_N} \int dX^+ dX^- \alpha r \bigg ( 2r \alpha^2 (\alpha^2-1) \left(P_{--+-} + P_{---+} \right)+ \nonumber\\  &\alpha^4 (P_{+-+-}+P_{-++-} + P_{+--+} + P_{-+-+} )+  \nonumber\\ &  4 r (\alpha^2-1)^2 P_{----} + 2 r \alpha^2 (\alpha^2-1) ( P_{+---} + P_{-+--} ) \bigg ).
\end{flalign}

Note that the components of the tensor $P_{\mu \nu \xi \eta}$ are non-negative functions of the degrees of the freedom such as central charge and the Kac-Moody level parameter. From the relation \ref{infmetricWADS}, one could note that the information metric for the warped AdS is an order of $\alpha^4$. As the effective entropy $S_{\text{eff} } (\Sigma_u)$ is proportional to $G_{uu}^{(B)}$, then we deduce that for the WAdS geometry also we have the relation $S_{\text{eff}} (\Sigma_u) \propto  \alpha^4 $. Therefore, it is a monotonically increasing function of $\nu$, but it becomes saturated at very big deformation parameters $\nu$. Its plot is shown in the right part of figure \ref{TNSEFF}.

Later, the implications of quantum estimation theory such as Cramer-Rao bound for the WAdS/WCFT could be studied.

\subsection{Building quantum circuits from Kac-Moody symmetry gates}\label{sec:WCFTgatesKac}

In this section, similar to the work of \cite{Caputa:2018kdj}, we can build the quantum circuits for the Kac-Moody algebra.

First, taking $T(x^-)$ and $P(x^-)$ as the usual local operators on the plane,  one could define the following operators \cite{Detournay:2012pc}

\begin{gather}
T_\zeta=-\frac{1}{2\pi}\int dx^- \zeta(x^-) T(x^-), \ \ \ \ \ P_\chi=-\frac{1}{2\pi} \int dx^- \chi(x^-)P(x^-).
\end{gather}

Here the right moving modes are associated with $x^-$ and left moving with $x^+$. Then, taking the coordinate transformation from $x^-$ to $\phi$, or from the Lorentzian plane to Lorentzian cylinder, in the following form
\begin{gather}
x^- = e^{i \phi},  \ \ \ \ x^+=t+ 2\alpha \phi,
\end{gather}
one would get
\begin{gather}
P^\alpha(\phi) = i x^- P(x^-)-k \alpha, \nonumber\\
T^\alpha(\phi)= - {x^-}^2 T(x^-)+\frac{c}{24}+ i 2 \alpha x^- P(x^-) - k \alpha^2.
\end{gather}

Then, the modes of the algebra on the cylinder could be written as
\begin{gather}
P_n^\alpha=- \frac{1}{2\pi} \int d\phi P^\alpha(\phi) e^{i n \phi}, \ \ \ \ \ \ L_n^\alpha=-\frac{1}{2\pi} \int d\phi T^\alpha(\phi) e^{i n \phi},
\end{gather}
which in terms of the original modes of Kac-Moody would be written as
\begin{gather}
P_n^\alpha=P_n+ k\alpha \delta_n, \ \ \ \ \ L_n^\alpha=L_n+2 \alpha P_n+(k \alpha^2- \frac{c}{24}) \delta_n. 
\end{gather}

Here $\alpha$ is an arbitrary \textit{tilt}.

The operators $L_n$ and $P_n$ of the Virasoro-Kac-Moody algebra would satisfy the following commutator relations
\begin{flalign}
[L_n, L_m]  \ &=  \ (n-m) L_{n+m}+\frac{c}{12} n (n-1) (n+1) \delta_{n+m} \nonumber\\
[P_n, P_m] \ &= \ \frac{k}{2} n \delta_{n+m}\nonumber\\
[L_n, P_m]\ &= \ -m P_{m+n}
\end{flalign}

Now similar to \cite{Caputa:2018kdj}, we could build the Virasoro$+$Kac-Moody symmetry gates as
\begin{gather}
U(\tau)= \cev{\mathcal{P}} \text{exp} \Big [ \int_0^\zeta Q[\zeta'] d\zeta'+ \int_0^\chi Q[\chi'] d\chi' \Big],
\end{gather}
where 
\begin{gather}
L_n= Q [\zeta_n], \ \ \ \ \ \ P_n= Q[\chi_n],
\end{gather}
are the associated charges of Virasoro-Kac-Moody $U(1)$ algebra, where the central extensions are \cite{Detournay:2012pc}
\begin{gather}
c=\frac{5\nu^2+3}{\nu (\nu^2+3)},  \ \ \ \ \ k=-\frac{\nu^2+3}{6\nu}.
\end{gather}

Here the instantaneous gate could be defined as
\begin{gather}
Q_T(\tau)= \int_0^{2\pi} \frac{ d\phi}{2\pi} \epsilon(\tau, \phi) T(\phi), \nonumber\\
Q_P(\tau)= \int_0^{2\pi} \frac{d\phi}{2\pi} \epsilon(\tau,\phi) P(\phi),
\end{gather}

where
\begin{gather}
T(\phi)=\sum_{n \in \mathbb{Z}} \Big( L_n+ 2\alpha P_n + (k \alpha^2 - \frac{c}{24}) \delta_n \Big) e^{- i n  \phi}, \nonumber\\
P(\phi) = \sum_{n \in \mathbb{Z}} ( P_n+ k \alpha \delta_n) e^{- i n \phi},
\end{gather}
and 
\begin{gather}
\epsilon (\tau, \phi)= \sum_{n \in \mathbb{Z}} \epsilon_n (\tau) e^{- i n \phi}.
\end{gather}

As mentioned in \cite{Detournay:2012pc}, the transformation 
\begin{gather}
P_n^\alpha=P_n+ k \alpha \delta_n, \ \ \ \ \ \ \ L_n^\alpha=L_n+ 2 \alpha P_n+(k \alpha^2-\frac{c}{24}) \delta_n,
\end{gather}
is the usual shift proportional to the central charge for an exponential mapping where its cost function $\mathcal{F}$ has been calculated in \cite{Caputa:2018kdj}, \textbf{combined} with a spectral flow transformation which would be given by that tilt parameter $\alpha$ which will be modeled by our chiraleon gate.

This gate could also be interpreted as generating anomaly in the system. This could be understood by considering the following coordinate transformation on the cylinder, as in \cite{Detournay:2012pc},
\begin{gather}
\phi=\frac{\phi'}{\lambda}, \ \ \ \ \ \ t=t'+2 \frac{\gamma}{\lambda} \phi'.
\end{gather}

As mentioned in \cite{Detournay:2012pc}, $T(x^-)$ is the generator for the coordinate transformations in $x^-$ and $P(x^-)$ is the generator for the gauge transformation in the gauge bundle which is parametrized by $x^+$. If similarly we assume the following functions for the coordinate transformations 
\begin{gather}
x^-=f(w^-), \ \ \ \ \ \ \ \ \ x^+=w^++g(w^-),
\end{gather}
then the general relations for the infinitesimal transformations, \cite{Detournay:2012pc},
\begin{flalign}
P'(w^-) \ &= \ \frac{\partial x^-}{\partial w^-} \Big [ P(x^-)+\frac{k}{2} \frac{\partial w^+}{\partial x^-} \Big ], \nonumber\\
T'(w^-) \ &= \ \Big( \frac{\partial x^-}{\partial w^-}  \Big)^2 \Bigg [T(x^-) -\frac{c}{12} \Bigg \lbrace \frac{ \frac{\partial^3 w^-}{\partial {x^-}^3}  } {\frac{\partial w^-}{\partial x^-} } -\frac{3}{2} \Bigg(\frac{ \frac{\partial^2 w^-}{\partial {x^-}^2 }  } { \frac{\partial w^- }{\partial x^-}   }  \Bigg)^2  \Bigg \rbrace \nonumber\\ &
+\frac{\partial x^-}{\partial w^-} \frac{\partial x^+}{\partial w^-} P(x^-)-\frac{k}{4} \Bigg ( \frac{\partial x^+}{\partial w^-} \Bigg)^2,
\end{flalign}
could be simplified as 
\begin{flalign}
T'(w^-)  &= f'^2 \Big[ T(x')-\frac{c}{12} \Big \{  \frac{f''-f' f'''}{f'}+\frac{3}{2} f''  \Big \} \Big] + f' g' P(x^-)-\frac{k}{4} g'^2,\nonumber\\
P'(w^-)  &= f' P(x^-)-\frac{k}{2} g',
\end{flalign}
where $f'= \frac{\partial f (w^-)}{\partial w^-}$ and $g'=\frac{\partial g(w')}{\partial x^-}$.

This is the corresponding equation $11$ in \cite{Caputa:2018kdj} where instead of the case of usual CFT, it works now for the warped CFT case. 

Also, there would be the following relations for the generator of translations,
\begin{gather}
Q[\partial_{t'}] = Q[\partial_t]+k \gamma, \ \ \ \ Q[\partial_{\phi'}]=\frac{1}{\lambda},
\end{gather}
which have been interpreted as constructing an anomalous term being generated by the new chiraleon gate.

Similar to \cite{Caputa:2018kdj}, the cost functions for the gate generating the ordinary transformation of the partial derivate and also the gate creating the anomaly, our chiraloen or \textit{"anomalon"} could be calculated as in \cite{Caputa:2018kdj}. First, we could find the cost functions in the following way
\begin{gather}
\mathcal{F}_1  \equiv \big | \text{Tr} \Big(\rho  \big (Q_T+Q_P  \big) \Big) \big | =  \big | \text{Tr} \Big(\rho_0 ( \tilde{Q}_T+\tilde{Q}_P)    \Big)\Big |, \nonumber\\
\mathcal{F}_2 \equiv \sqrt{-\text{Tr} \big(\rho (Q_T+Q_P)^2  \big) } =\sqrt{- \text{Tr} \big (\rho_0 ({{\tilde{Q}}^2_T +{\tilde{Q}}^2_P)}  \big)},
\end{gather}
where
\begin{gather}
\rho \equiv U \rho_0 U^\dagger, \ \ \ \ \ \tilde{Q}_T=U^\dagger Q_T U, \ \ \ \ \tilde{Q}_P=U^\dagger Q_P U.
\end{gather}

So we have
\begin{flalign}
\tilde{Q}_T \ &= \int_0^{2\pi} \frac{dx^+}{2\pi}  \epsilon(x^+,x^-) \Bigg( f'^2 \Big[ T(x')-\frac{c}{12} \Big \{ \frac{f''-f' f'''}{f'}+\frac{3}{2} f'' \Big \} \Big] +f' g' P(x')-\frac{k}{4} g'^2 \Bigg), \nonumber\\
\tilde{Q}_P \ &= \int_0^{2\pi} \frac{d\sigma}{2\pi} \epsilon(x^+,x^-) \Big( f' P(x^-) -\frac{k}{2} g' \Big).
\end{flalign}

We can insert the expectation values of $T(x')$ and $P(x')$ in the above relations as
\begin{gather}
T(x') \to  \big | k \alpha^2-\frac{c}{24} \big |, \ \ \ \ \ \ P(x') \to \big | k\alpha \big |.
\end{gather}

The last thing to do is to relate the velocities $\epsilon(x^+, x^-)$ to the path in the group manifold using the symmetries of Kac-Moody algebra, which would lead to the result, $ \epsilon(x^+, x^-)=\frac{1}{ f' g'}$.

At the end, similar to \cite{Caputa:2018kdj}, the complexity action could be computed using the cost functions $\mathcal{F}_1$ or $\mathcal{F}_2$, which for the warped CFT becomes
\begin{flalign}
\mathcal{C} (\tau)  &= \int^\tau \mathcal{F} (H (\tau ')) d\tau  \nonumber\\ & =
\frac{1}{2\pi} \int_0^{x^-} d {\tilde{x}}^-  \int_0^{2\pi} d {\tilde{x}}^+  \Bigg ( \frac{f'}{g'} \big[ k\alpha^2-\frac{c}{24}-\frac{c}{12} \big \{ \frac{f''-f' f'''}{f'}+\frac{3}{2} f'' \big \} \big]+k \alpha -\frac{k}{4} \frac{g'}{f'} \Bigg) \nonumber\\  & + 
\frac{k}{2\pi} \int_0^{x^-} d {\tilde{x}}^-  \int_0^{2\pi} d {\tilde{x}}^+  \Big ( \frac{\alpha}{g'}-\frac{1}{2f'} \Big).
\end{flalign}

In the gravity side, this result could also be derived using Polyakov action in the chiral gauge. So as we pointed out before, for calculating complexity, Polyakov action should be considered as the fundamental theory rather than the Liouville gravity.

The equivalent connection between Polyakov action and the coadjoint orbit action of the Virasoro group has been shown in \cite{Alekseev:2018ful}, and the connection between quantum complexity and coadjoint orbit action in \cite{Caputa:2018kdj}. So, here, we have emphasized that complexity is not necessarily equivalent to coadjoint orbit action of the Virasoro group. This is only true if one chooses the conformal gauge fixing for the Polyakov action. If one chooses the chiral gauge, then complexity would be related to coadjoint orbit action of the Virasoro$+$ Kac-Moody group \cite{Barnich:2017jgw}.

Then, for the case of warped geometries, the results of path-integral complexity could be compared with those of holographic calculations using the $\text{complexity}=\text{action}$ or $\text{complexity}=\text{volume}$ \cite{Ghodrati:2017roz, Auzzi:2019fnp, Auzzi:2018pbc}.

\subsection{Kinematic Space for Warped CFT}\label{sec:WarpedKinematic}

Kinematic space is an auxiliary Lorentzian geometry where the components of its metric is being defined using conditional mutual information \cite{Czech:2015xna}. This concept would help to understand the interplay between information on the boundary CFT and the bulk geometry better.  For holographic CFTs where Ryu-Takayanagi (RT) prescription works, kinematic space would be just the space of bulk geodesics. The lengths of the bulk curves could then be calculated by the volumes in the kinematic space.

The kinematic space in general is related to the number of isometry gates in the tensor network and at least in the first order of approximation, it could be constructed by calculating the Crofton form \cite{Balasubramanian:2013lsa,Czech:2015kbp,Czech:2015qta}

\begin{gather}
\omega(\theta, \vartheta)=\frac{\partial ^2 S(u,v) } {\partial u \partial v} du \wedge dv,
\end{gather}
where
\begin{gather}
u= \theta-\vartheta, \ \ \ \ \ \ v=\theta+\vartheta.
\end{gather}

Now we would like to study the properties of this space for the warped CFTs and check how the warping factors and also the symmetry breaking change its properties. In this case, the geodesics are located in a space were they asymptote to a warped $\text{AdS}_3$ background. 

For the case of  the static slice of AdS, the kinematic space would be a two-dimensional de-Sitter space. For warped CFTs, one could imagine that its kinematic space which is the space of intervals on a constant time slice of a given two-dimensional ``warped CFT", would then be a two-dimensional ``deformed de-Sitter" space.

If one considers an $\text{AdS}_3$ background which is expressed as a real-line or a circle fibration over a Lorenzian $\text{AdS}_2$ base space, then these geometries could certainly get deformed with a warped factor into the warped $\text{AdS}_3$ which then affects the kinematic space.

In addition, the RT-like prescription for the warped AdS also has a modified version.These structures have been discussed in \cite{Anninos:2013nja,Song:2016gtd,Song:2016pwx}, which we will use in our calculations later.

As mentioned in \cite{Czech:2015qta}, there would be a relation between the sign of the Crofton form and the strong subadditivity leading to the result
\begin{gather}
I(A,C,B)= S(AB)+S(BC)-S(B)-S(ABC) \ge 0, \nonumber\\
\approx \frac{\partial ^2 S(u,v) }{ \partial u \partial v} du dv \ge 0.
\end{gather}

The volume of the kinematic space for the case of MERA is determined by the number of these isometries. This of course still could be true for the case of deformed MERA since we could write
\begin{gather}
ds^2_{\text{MERA}}=I (\Delta u, \Delta v \big | B).
\end{gather}

\begin{figure}[ht!]
\centering
\includegraphics[width=0.5\textwidth]{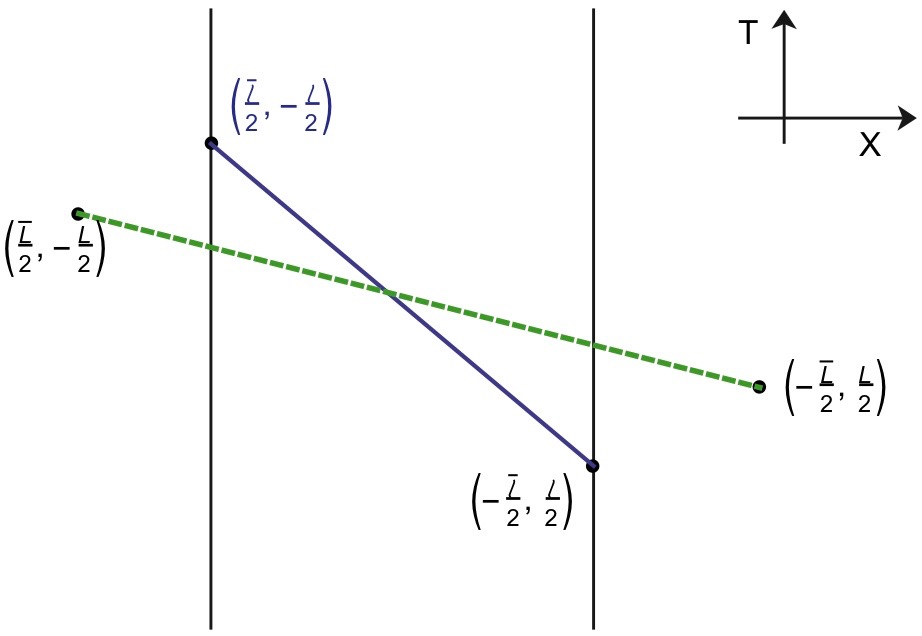}
\caption{The interval setup $\mathcal{D}$ considered in \cite{Castro:2015csg} which shows the domain covered by the coordinates $(t,x)$ relative to $(T,X)$. We implement this plot to construct the structure of kinematic space for WCFTs.}
\label{wapredDomain}
\end{figure}

Also, similar to the work of \cite{Callebaut:2018xfu}, we conjecture that the dynamics of the kinematic space for warped CFTs would be the dynamics of the two-dimensional chiral Liouville gravity, which would be the corresponding deformed version of Jackiw-Teitelboim (JT) gravity. In this case however, as shown in \cite{Jensen:2017tnb}, the throats of $\text{WAdS}_3$ with a finite volume of spatial circle would not completely decouple from the rest of the geometry. Similar to the CFT case though, the chiral Liouville stress tensor for the entanglement would be given by the vacuum expectation value of WCFT stress tensors evaluated at the interval endpoints.

Now, for calculating the entanglement entropy and then the Crofton forms, we could use the procedures done in \cite{Castro:2015csg} and their coordinate transformation shown in figure \ref{wapredDomain} and then we could  construct the kinematic space for warped CFTs.

First, we consider an interval $\mathcal{D}$ on a background which has the cylinder geometry described by coordinates $(T,X)$ where $T$ is related to $U(1)$ axis and $X$ to the $SL(2, \mathbb{R})$ symmetry. The identification of the cylinder would be $(T,X) \sim (T+\bar{L},X-L)$. The interval inside this cylinder would be taken similar to the one in \cite{Castro:2015csg} as
\begin{gather}
\mathcal{D}:  \  (T,X) \in  \Big [ ( \frac{ \bar{\ell} }{2},-\frac{\ell}{2}),(-\frac{\bar{\ell} }{2}, \frac{\ell}{2}) \Big] .
\end{gather}

After mapping to the appropriate coordinate $(t,x)$, the interval would become
\begin{gather}
(t,x) \in \Big [ (\frac{\bar{\kappa}}{2\pi} \zeta-\frac{\ell}{2} \frac{\bar{L}}{L}+\frac{\bar{\ell} }{2}, -\frac{\kappa}{2\pi}\zeta), (-\frac{\bar{\kappa} }{2\pi} \zeta+\frac{\ell}{2} \frac{\bar{L} }{L}-\frac{\bar{\ell} }{2}, \frac{\kappa}{2\pi} \zeta) \Big ],
\end{gather}
where
\begin{gather}
\zeta= \log \Big ( \frac{L}{\pi \epsilon} \sin \frac{\pi \ell}{L} \Big)+ O(\epsilon). 
\end{gather}

Then, the relation for the entanglement entropy of warped CFTs mentioned before, \cite{Castro:2015csg}, in the form
\begin{gather}\label{eq:warpedEE}
S_{EE}= i P_0^{\text{vac}} \ell \Big ( \frac{\bar{L}}{L}-\frac{\bar{\ell}}{\ell} \Big ) -4 L_0^{\text{vac} } \log \Big ( \frac{L}{\pi \epsilon}  \sin \frac{ \pi \ell}{L} \Big ),
\end{gather}
could be applied

\begin{figure}[ht!]
\centering
\includegraphics[width=0.25\textwidth]{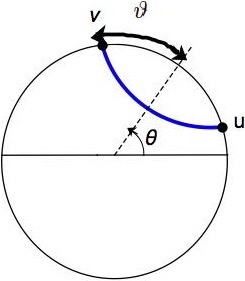} \hspace{1cm} 
\includegraphics[width=0.35\textwidth]{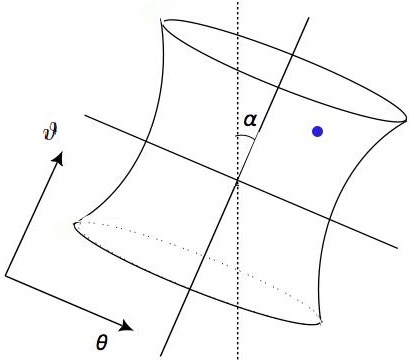}
\caption{ The geodesic and the corresponding point in the kinematic space are shown in blue color. Also, the angles $\vartheta$, $\theta$ and $\alpha$ in relation \ref{eq:angleKinematic} for the warped $\text{AdS}_3$ case are shown. However, the tilt angle $\alpha$ introduced in relation \ref{eq:modifiedwarpedEE} which should be fixed by the theory and is ``not" an arbitrary parameter of warped AdS backgrounds, would not necessarily have the exact geometrical meaning shown here.}\label{fig:warpedKinematicangle}
\end{figure}

Similar to the CFT case, we take $L=2\pi \ell_{\text{WCFT}}$, $\ell=\ell_{\text{WCFT}} (v-u)$ and also $u=\theta-\vartheta$, $v=\theta+\vartheta$.  All the angles for the warped AdS case are shown in figure \ref{fig:warpedKinematicangle}. 

Taking the corresponding relations for the conjugate ones, the above relation could be written as
\begin{gather}
S_{EE}=i P_0^{\text{vac}} \Big ( \bar{\ell}_{\text{WCFT}} (v-u)-\ell_{\text{WCFT}} (\bar{v}-\bar{u}) \Big)-4 L_0^{\text{vac}} \log \Big( \frac{2 \ell_{\text{WCFT}}}{\epsilon} \sin \frac{v-u}{2}\Big).
\end{gather}

Using above relation to derive the Crofton forms, one would get the CFT corresponding relations as
\begin{gather}
\omega= \frac{\partial^2 S_{EE} }{\partial u \partial v} du \wedge dv=-\frac{1}{2} \partial_\vartheta^2 S d\theta \wedge d\vartheta=\frac{4 L_0 }{\sin^2( \vartheta)} d\theta \wedge d\vartheta, \ \ \ \  \ \bar{\omega}=0,
\end{gather}
which does not seem to be a plausible result.

However, if we use the corrected, ``modified" result for the entanglement entropy of WCFTs found in \cite{Song:2016gtd,Apolo:2018oqv} as
\begin{gather}\label{eq:modifiedwarpedEE}
S_{EE}= i P_0^{\text{vac}} \ell  \left( \frac{\bar{L}-\alpha}{L}-\frac{\bar{\ell} }{\ell} \right)+ \left(-i \frac{\alpha}{\pi} P_0^{\text{vac}} -4 L_0^{\text{vac}} \right) \log \left( \frac{L}{\pi \epsilon} \sin \frac{\pi \ell}{L} \right),
\end{gather}
which for the case of zero tilt angle $\alpha=0$ leads to the previous result \ref{eq:warpedEE}, then for the Crofton form, we could get the correct and reasonable result. Changing variables as before simply leads to 
\begin{gather}\label{eq:angleKinematic}
\omega= \frac{\partial^2 S_{EE} }{\partial u \partial v} du \wedge dv=-\frac{1}{2} \partial_\vartheta^2 S d\theta \wedge d\vartheta=\frac{ \frac{i \alpha}{\pi} P_0^{\text{vac}}+ 4 L_0 }{\sin^2 (\vartheta)} d\theta \wedge d\vartheta, \ \ \  \ \bar{\omega}=0.
\end{gather}

The difference in the volume form now is the term $\frac{i \alpha P_0^{\text{vac}}}{ \pi \sin^2 \vartheta}$ which is the gap in the ``density of geodesics" between CFTs and warped CFTs.

 As mentioned in \cite{Song:2016gtd}, $\alpha$ is not an arbitrary parameter and actually is being fixed by the theory. This statement corresponds to the statement that the entropy is also invariant under the ``warped conformal transformations".

Our results could then help to reveal the nature of gates in the tensor network structures of warped CFTs.  The density of ``chiraleon gates" in such tensor networks would be proportional to this additional term which depends on the $U(1)$ charge and the tilt angle $\alpha$. This tilt parameter $\alpha$ then will produce the tilt angle in the tensor network structure shown in figures \ref{fig:tensorWarpedParameters} and \ref{fig:warpedKinematicangle}. This result could also give further evidences that the ``modified" relation for the entanglement entropy of WCFTs is indeed the correct one. The modified term actually shows its signature in the volume form, geodesic density and tensor network structure.

In \cite{Song:2017czq}, it was mentioned that $\alpha$ is proportional to the slope of the thermal identification after the modular transformation and taking a finite value for $\alpha$ corresponds to the slow rotating limit of \cite{Detournay:2012pc}
\begin{gather}
c \gg \beta \Omega, \ \ \ \ \ \ \ \frac{\Delta_{\text{gap} } }{\beta \Omega} \gg 1.
\end{gather}

In the above relations, $\Delta_{\text{gap}}$ is the dimension of $L_0$ where the theory starts to get a large number of operators. From these statements then, one could deduce that inserting the ``chiraleon" gates in the tensor network would lead to such slow-rotating regimes.

\section{Cost functions, Polyakov Action and Generalizations}\label{sec:Polymodes}

The exact relations between circuit complexity, quantum gates, tensor network models and the actual properties of string theory deserve further studies. By comparing various results from Polyakov action in conformal versus chiral gauge, we could gain several results here in this direction. 

Interesting factors such as choosing gauge, boundary conditions and its symmetry algebra would then affect the choice of cost functions and quantum gates for calculating circuit complexity. For instance the behavior of the boundary modes, i.e, symmetry of the left and right-moving modes in the conformal case, versus the only left moving modes of chiral gauge, would affect the nature of quantum gates to be chosen for the TN models of these cases. Therefore, the extension of Liouville action as the cost function is inevitable.

To see this point better, we review the geometric definition of complexity introduced in \cite{Jefferson:2017sdb, Chapman:2017rqy}. There, the circuit is modeled by an operator of the form
\begin{gather}
V= \mathcal{P} \text{exp} \left ( - \int_0 ^\lambda d \kappa \sum_I Y_I(\kappa) \mathcal{O}_I \right),
\end{gather}
where $\mathcal{O}_I$ are the elementary gates coming along each other in a sequence and $Y^I$ controls the insertion of these gates at each layer, (which could be considered as a velocity along the path $\kappa$). The connection between the nature of these elementary gates $\mathcal{O}_I$ and the properties of strings, such as tension $T$ or mass could be studied. The gates are related to the components of the energy momentum tensor, and if one considers more generalized form of Liouville, for instance the Polyakov action, one could find also connections between the string tension $T$, or the Regge slope $\alpha'$ and nature of these quantum gates.

In fact, the cost function $\mathcal{D}$ \cite{Camargo:2019isp}, would be related to the dynamics of the string, and the way the end-points of these strings would be coupled to any boundary or D-brane. For instance, taking the $L^2$-norm $\mathcal{D}_2 =\int_0^\lambda d\mathcal{\kappa} \sqrt{\sum_{IJ} \eta_{IJ} Y^I Y^J }$ for calculating complexity would correspond to taking the Nambu-Goto action describing the dynamics of the strings. In \cite{Camargo:2019isp}, the Liouville action has been proposed to be a good cost function which also match with result of \cite{Miyaji:2016mxg}. Then,  here we conjectured that, as we have an analogue warped Weyl invariance for WCFTs, for these theories, the chiral Liouville action would be a suitable cost function. This argument could then be extended to the Polyakov action and its various generalization discussed below.

Let's look at the Polyakov action more closely,
\begin{gather}
S=\frac{T}{2} \int d^2 \sigma \sqrt{-h} h^{ab} g_{\mu \nu} (X) \partial_a X^\mu (\sigma) \partial_b X^\nu (\sigma).
\end{gather}

Here $T$ is the string tension, $g_{\mu \nu}$ is the metric of the target manifold and $h_{ab}$ is the worldsheet metric.

Remember the connection between the kinematic and potential terms of Liouville action with the number of isometries and disentanglers of MERA network. The unitaries of the system which are proportional to the potential term could execute a force on the boundary modes of the system which would be related to the tension of the string $T_s$, or other quantities of the model such as $\mu$ (the mass of graviton) for the case of massive gravity. The disentanglers then, by removing the entanglement entropy between the virtual particles would create a kinetic energy, and consequently the number of isometries or disentanglers would be proportional to the kinetic term in the Liouville action. So for all these cases, an important parameter would be the tension of the open strings.

\begin{figure}[ht!]
\centering
\includegraphics[width=0.3\textwidth]{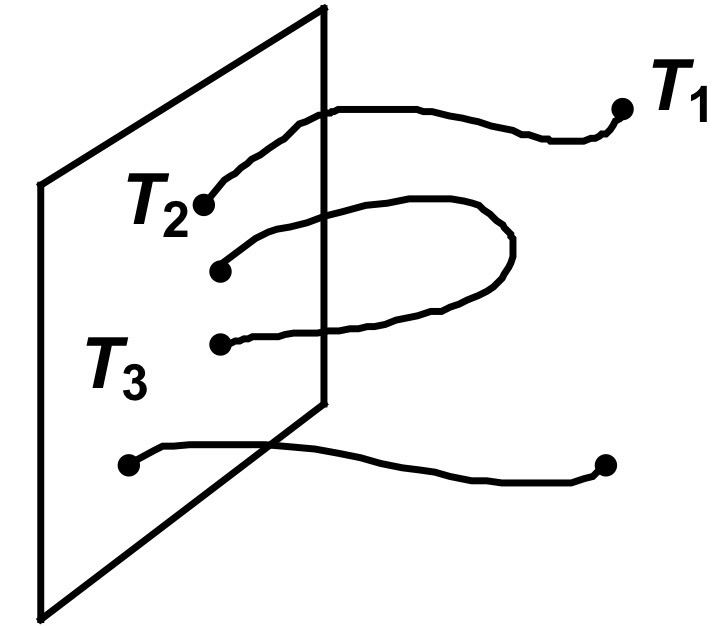}
\caption{ The different tensions in the system, $T_1$ is the tension of the string, $T_2$ is the tension between the string and the worldsheet and $T_3$ is the brane tension itself. }\label{fig:Stringtensions}
\end{figure}

To consider a more generalized form, one could imagine that each of these coupling parameters in the action corresponds to a different tension, see figures \ref{fig:Stringtensions}, which then leads to the generalized form of Polyakov action where there are various $T_i$ parameters. This action then could offer a more inclusive cost function for calculating the circuit complexity and for constructing various tensor network models. 

It worths to note here that the non-linear sigma models, such as $O(n)$ model as another example, have many interesting features which could give a better picture. For instance the non-trivial renormalization-group flow \textit{fixed point} of the $O(n)$ symmetric model contains a critical point separating the ordered phase from the disordered one. This phase transition could also be captured by the tensor network models. The specific nature of the quantum gates and the string tension $T_1$, brane tension $T_2$, and the tension between string and the brane $T_3$ would affect this phase transition then. This model could even be experimentally closer to the results of $O(n)$ models describing the Heisenberg ferromagnets.

One could even go beyond that and think of more generalized actions in string theory, in order to find the suitable cost function for the circuit complexity. For instance one action, could be written in the form of sigma model action for \textit{``$p$-adic strings"} which has been proposed by Zabrodin  \cite{Huang:2019nog,Zabrodin:2017ryf, Gubser:2017pyx} as
\begin{gather}
S \sim \sum_{\left< i j \right> \in E(T_p)}\eta_{ab} (X_i^a-X_j^a) (X_i^b-X_j^b),
\end{gather}
where $\eta_{ab}$ is target space Minkowski metric, the worldsheet is Bruhat-Tits tree $T_p$, and $\left < ij \right>$ is the edge between vertices $i$ and $j$.  The generalized version of AdS/CFT using this Bruhat-Tits tree would be more suitable for constructing tensor network and so it could be more compatible with the results of quantum error correction which was introduced in \cite{Gubser:2016guj,Gubser:2017tsi}. The tensor network for the case of $\text{AdS}_3/\text{CFT}_2$, using p-adic strings was constructed in \cite{Heydeman:2016ldy}. 

Then, in \cite{Huang:2019nog}, the authors generalized the Zabrodin action which is equivalent to Polyakov action, to the case where the target space is changed from a flat geometry to a curved one.  The proposed action would be in the form
\begin{gather}
S=\sum_{\left< i j \right> \in E(T_p)} \frac{d^2 (X_i, X_j)}{V a^2_{\left< i j \right> } }, 
\end{gather}
where $d(X_i,X_j)$ is the target space distance between the two points $X_i$ and $X_j$ and $a_{\left< i j \right>}$ is the length of edge $\left< i j \right>$ in the tree, and $V$ is the degree of any vertex. This action could be one of the most generalized form to be considered as the cost function for circuit complexity.

Adding nonlinear interactions such as flavor-chiral anomalies (which would lead to Wess-Zumino-Witten model) and extending geometry to include torsion or considering the cobordism classes and its relations to complexity and swampland would be other non-trivial directions to generalize these arguments, \cite{McNamara:2019rup,Wan:2019fxh}.

 \section{Discussion}\label{sec:Discuss}
 
 In this work we studied the optimization path integral complexity for the case of warped conformal field theory in the setup of $\text{WAdS}_3/\text{WCFT}_2$. For this computation we proposed the chiral Liouville action as the suitable cost function which then could lead to the slice of a warped geometry. The RG equations where two cut offs would be needed, one for each field in the metric, have been presented. Then, the RG equations for two special cases where the Lagrangian of WCFTs have been directly obtained where discussed. Those Lagrangians have been written for a scalar field action and a Weyl fermion action called Hofman/Rollier theory.
 
We then looked at the problem of emergence of warped AdS geometry from the chiral warped CFT from the other side of the story.  So, we started from the warped $\text{AdS}_3$ and by calculating the action of TMG as an example which has the solution of warped $\text{AdS}_3$ (in addition to other solutions such warped BTZ black holes), we derived three warped Liouville actions from the \textit{spacelike, timelike} and \textit{null} warped $\text{AdS}_3$ metric. We showed that the specific boundary parts which arise from the topological terms are consistent with the final emergent spacetimes one would expect. It would be interesting then to repeat these calculations for other gravity theories which contain warped $\text{AdS}_3$ as their solutions such as NMG, and then derive new forms of deformed Liouville actions. Also, the specific forms of Jacobians and coordinate transformations using various boundary cut off profiles should be studied further.

Then, we implemented various holographic tools that have already been constructed for the case of $\text{AdS}_3/\text{CFT}_2$, in the case of $\text{WAdS}_3/\text{WCFT}_2$. These include a deformed version of MERA for the tensor network of warped CFTs. Specifically, based on the symmetry group of the theory, we introduced a new kind of gate one would need to construct the desired TN model, where we dubbed  \textit{``chiraleon"}. Furthermore, we discussed some subtleties of putting fermions on a lattice which are relevant to our discussions.  

We also discussed the entanglement structure of WCFTs which include a twist and discussed how it affects the emergent warped AdS geometry.  

The surface/state correspondence for the case of warped AdS and chiral theories has also been discussed where we showed the relationship between the deformation parameter of the geometry and the entanglement entropy.  Also, a form of Fisher information metric for the warped cases has been proposed there.

Next, similar to the approach of \cite{Caputa:2018kdj} done for the Virasoro group, the quantum circuits model for the Kac-Moody group has been proposed and then using them and also a specific cost function, the complexity for WCFTs in terms of the parameters of the algebra such as central charge, Kac-Moody level and symmetry transformation functions, has been derived. Finally, the structure of kinematic space for the case warped AdS/warped CFTs has been analyzed.

In the final section, we discussed the various extensions of Liouville action (in the conformal gauge) as the cost function for complexity to the case of chiral Liouville for the chiral gauge and then to Polyakov action and even its generalizations such as Zabrodin action or p-adic strings which could be used to construct tensor network models. The exact connection between each of these actions, the equations of motion, calculations of entanglement entropy and complexity and the connections with the tensor network structures deserve further investigations.

So in this paper, we emphasized the role of gauge fixing in constructing tensor network and calculating complexity. Recently, the question of gauge fixing for tensor networks has also been studied further in \cite{PhysRevB.98.085155}. There, it has been mentioned that the correct gauge fixing could in fact help to optimize the iteration process of algorithms. 

In light of many recent developments, there would be various directions to generalize and extend this work which here we mention to a few ideas. 

In \cite{Ghodrati:2019hnn}, the connection between entanglement of purification (EoP) and complexity of purification (CoP) has been studied. These studies could also be done for the case warped conformal field theories and check how the lack of dilation in one direction could change the EoP and CoP and their evolutions \cite{Zhou:2019jlh}. 

In \cite{Camargo:2019isp}, it was proposed that by including non-unitary transformations \cite{Agon:2018zso,Camargo:2018eof} in the Fubini-Study metric, one could define mixed state complexity. It would be interesting to see if such non-unitary transformations could also create a chiral state. Also, one could connect the ``Fubini-Study metric" to chiral Liouville gravity, Polyakov action or its generalizations.

In \cite{Callebaut:2018xfu}, the gravitational dynamics of kinematic space have been discussed and there it has been argued that it could be described by the ``Jackiw-Teitelboim" gravity theory. The relationship between the modular Hamiltonian and the dilaton in the gravity model which underlies the kinematic space construction has been discussed. The complete set of equations for the connections between entanglement entropy $S$ and modular Hamiltonian $H_{\textit{mod}}$ have been presented where it has been called the \textit{kinematic space on-shell identities}.  These dynamical relations and similar equations for the case of warped AdS could also be derived.

The construction of \cite{Harlow:2019yfa} could specifically be useful in studying the Hamiltonian dynamics of WCFTs. The symplectic structure of warped CFT could be constructed which would help to understand the holographic dictionary for WAdS/WCFT better. The construction for TMG could be a nice example for applying their algorithm, since in their method one specifically could treat the important boundary terms of TMG in a more complete and efficient way which then could help to understand our derived complexity relations better.

In \cite{Franco-Rubio:2019nne, Zou:2019xbi}, the construction of cMERA for gauge theories has been discussed, specially both a massless and massive $U(1)$ gauge theory have been studied.  So it has been shown there that gauge invariance of a theory and \textit{``quasi-local"} character of the \textit{entangler} which generates the cMERA wave functional are indeed compatible with each other. This could be a good evidence that cMERA would also work for the case of warped conformal field theories.  Note that this compatibility could also work for the interacting or non-Abelian gauge theories.

Then, all of these studies could be repeated for other Lorentz-violating backgrounds such as Lifshtiz or hyperscaling violating geometries \cite{Ghodrati:2014spa}, and specifically using the already calculated results for the entanglement entropy of these backgrounds, in various works, one could check the effects of Lifshitz parameter $z$ or hyperscaling exponent $\theta$ on the structure of the tensor network, kinematic space, quantum gates of symmetries, cost functions, Fisher information metric, etc.

These studies could also be performed again for the case of $T \bar{T}$ and $J \bar{T}$ deformations. For instance, in \cite{Ota:2019yfe}, it has been found that the integrable $T \bar{T}$ deformation would decrease the degrees of freedom of the subsystem leading to a renormalization-like flow. So one could consider this deformation as the rescaling of the energy scale. The phase transitions in these systems have been studied as well. Similar calculations then for the case of warped CFTs or $J\bar{T}$ deformations could be performed. Recent progress on path-integral optimization for $T\bar{T}$ has been reported in \cite{Jafari:2019qns}.

The specific structure of casual cone of warped AdS could also be constructed and then the theory of minimal update proposal (MUP) and rayed MERA \cite{Czech:2016nxc} could be applied to the warped CFTs. The existence of the anomalous symmetry and chirality could have interesting consequences in this setup. 

Also, in \cite{Ghosh:2015iwa, Soni:2015yga}, some new aspects of entanglement entropy in ``gauge theories" have been studied. Using path integral complexity, the implications of these aspects of gauge theories for the complexity could also be analyzed. One, for instance, could check how the notion of path-integral complexity would be different for the Abelian versus non-Abelian gauge theories. One specifically could also check which extended definition could match better with the number of Bell pairs associated with the complexity of purification introduced in \cite{Ghodrati:2015rta}.

Additionally, the emergence of time from the density of unitary quantum gates and the emergence of gravitational force, or other kinds of forces such as electric or magnetic, from the quantum circuits in the setup similar to \cite{Takayanagi:2018pml} could be studied further.

\section*{Acknowledgement}

I thank Jian-Pin Wu for his help during the calculations of ``spin connections". I am grateful to Bartlomiej Czech, Wei Song, Tadashi Takayanagi and Kento Watanabe for insightful discussions. I also thank the Yukawa Institute for Theoretical Physics at Kyoto University where discussions during the workshop YITP-T-19-03 ``Quantum Information and String Theory 2019" were useful to complete this work.

 \medskip

\bibliography{warpedComplexityJHEP}
\bibliographystyle{JHEP}
\end{document}